\documentclass[journal]{IEEEtran}

\usepackage[utf8]{inputenc}
\usepackage{cite}
\usepackage{graphicx}

\usepackage{mhchem}

\usepackage{comment}
\usepackage{amsmath,amssymb,amsthm,mathtools,amsfonts}
\usepackage{bm}
\usepackage[linesnumbered, ruled]{algorithm2e}%
\usepackage{bbm}
\usepackage{url}
\usepackage{array}
\usepackage{hyperref}
\usepackage{enumerate}
\usepackage[squaren]{SIunits}
\usepackage{cases}
\usepackage{bbold}

\usepackage{xcolor}

\def\ci{\mathrm{CR}}
\def\im{\varphi}
\def\imm{\varphi_{3,4}}
\def\den{\xi}
\def\reg{\mathcal{R}}
\def\fov{\text{FOV}}
\def\patch{Z}
\def\covt{\text{Cov}}
\def\hesst{\mathcal{H}}
\def\hess{\mathbf{H}}
\def\of{\mathcal{O}}
\def\vart{\text{Var}}
\def\finp{\f^{\neg \patch}_{\im_{1,2}}}
\def\null{\mathrm{N}}
\def\inp{\mathrm{I}}
\def\pro{\mathrm{Pr}}%
\def\fspace{\mathrm{F}}
\def\fdim{h}

\def\v{\mathbf{v}}

\def\vb{\mathbf{b}}
\def\f{\mathbf{f}}
\def\inte{\phi}
\def\frand{\tilde{\mathbf{f}}}
\def\vrand{\tilde{\mathbf{v}}}
\def\interand{\tilde{\phi}}
\def\imrand{\tilde{\im}_{1,2}}
\def\brand{\tilde{\vb}}
\def\frandc{\frand |\im_{1,2}}
\def\map{\f^\star_{\im_{1,2}}}

\def\b{\mathbf{b}}
\def\v{\mathbf{v}}

\def\x{\mathbf{x}}

\def\reals{\mathbb{R}}

\def\naturals{\mathbb{N}}

\begin{document}
\bstctlcite{IEEEexample:BSTcontrol}

\title{How accurate is mechanobiology? \\ A statistical test of cell force}

\author{Aleix Boquet-Pujadas
\\\vspace*{-1pt}
}

\maketitle

\begin{abstract}
    Mechanobiology is gaining more and more traction as the fundamental role of physical forces in biological function becomes clearer. Forces at the microscale are often measured indirectly using inverse problems such as Traction Force Microscopy because biological experiments are hard to access with physical probes. In contrast with the experimental nature of %
biology and physics, 
these measurements do not come with error bars, confidence regions, or p-values. The aim of this manuscript is to publicize this issue and to propose a first step towards a remedy therefor in the form of a general reconstruction framework. We also show that this opens the door to hypothesis testing of seemingly abstract experimental questions.

\end{abstract}


Although long considered subordinate to molecular chemistry, physical forces are now recognized as fundamental to biological function at all scales~\cite{iskratsch_appreciating_2014,editorial_forces_2017}. 
They are necessary for tissue morphogenesis~\cite{mongera_fluid--solid_2018}, they influence cell metabolism~\cite{romani_extracellular_2019}, and they regulate transcription, for example, via nucleocytoplasmic transport or chromatin deformation~\cite{tajik_transcription_2016, andreu_mechanical_2022}.

Taking physical measurements at the microscale is challenging, especially because experimental setups are often difficult to access~\cite{marx_may_2019}. Methods that rely on physical probes, such as micropipette aspiration and atomic-force microscopy, are precise but invasive, while image-based methods can ``see'' through obstacles without interfering, albeit at the cost of increased computational complexity~\cite{boquet-pujadas_bioimage_2021}. 
The latter methods are becoming more prevalent as biological experiments grow increasingly intricate in pursuit of physiological relevance~\cite{boquet-pujadas_bioimage_2021}. In either case, an underlying model is needed because physical forces can only be measured (indirectly) through their effects on known materials. This relationship is captured by Newton's second law~\cite{roca-cusachs_quantifying_2017}. 

\begin{figure*}[!ht]
    \centering
    \includegraphics[width=\linewidth]{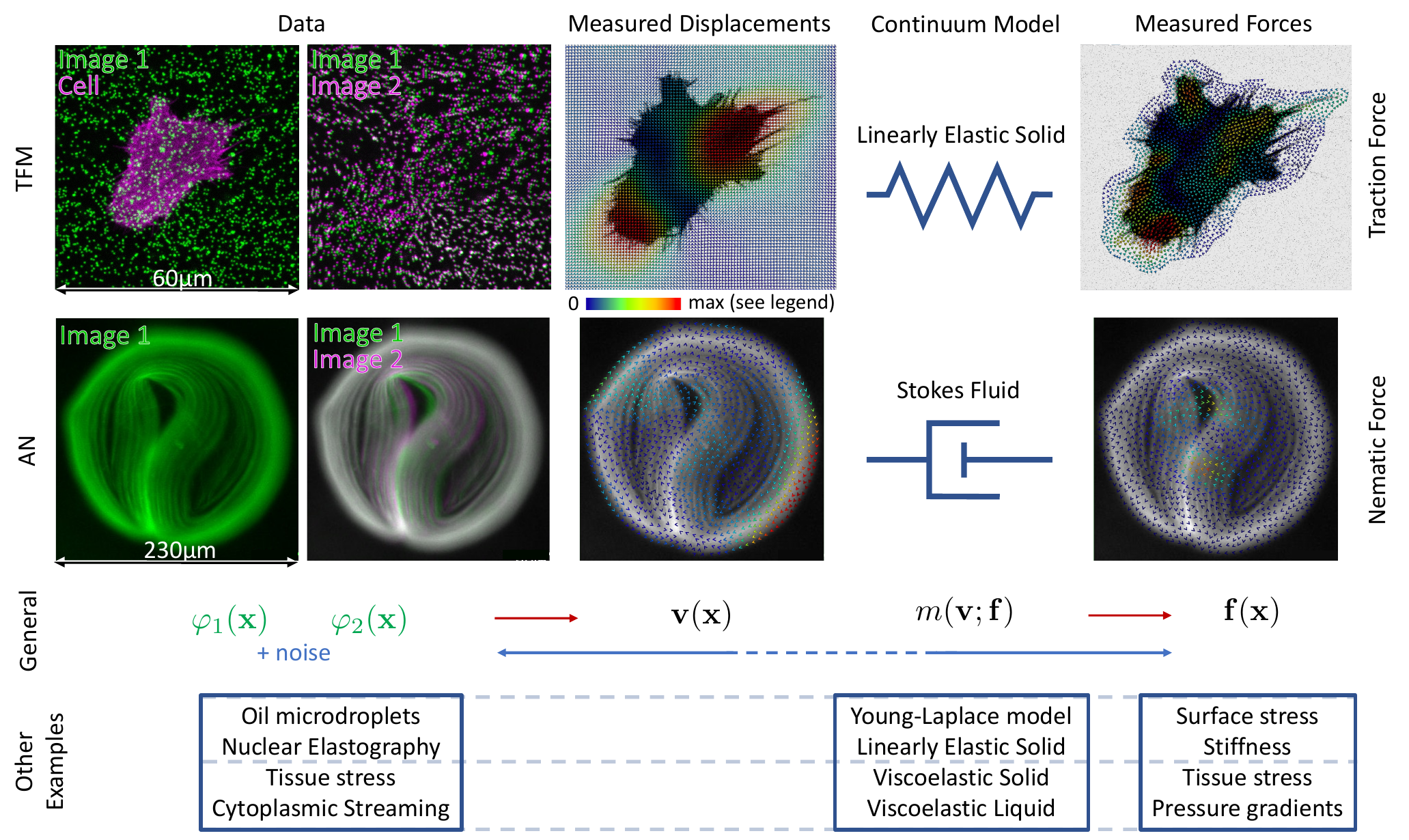}
    \caption{\textbf{A General Framework for Image-Based Mechanobiology.} \textbf{Top row}: application of the framework to Traction Force Microscopy (TFM). Left to right: one image (green) of the two required for TFM capturing the fluorescent beads in the substrate with the cell (magenta) overlaid; second image (green, substrate at rest) of the two with the first one overlaid (magenta, substrate under traction) to help with the visualization of movement; displacement measurements resulting from our framework applied to TFM (colorbar range \unit{0-2.1}{\micro \meter}); continuum model used for TFM; traction-force measurements resulting from our framework applied to TFM (range \unit{0-0.16}{\micro \meter^{-1}}). \textbf{Second row}: same as Top but for taking measurements in Active-Nematics (AN) systems. The colorbar ranges are \unit{0-2.9}{\micro \meter . \second^{-1}} for the velocity, and \unit{0-5.6\cdot 10^{-2}}{\micro \meter^{-1} . \second^{-1}} for the (relative) nematic forces. \textbf{Third row}: General mathematical formulation (see text) of the framework. The red arrows represent the two steps of classical techniques, e.g. for TFM. The blue arrows represent how our framework links the image data directly to the measurement of interest and creates a feedback loop while taking the noise into account. \textbf{Bottom row}: Examples of other image-based measuring techniques that can be reformulated into our framework using different systems and models (e.g., Zener or Jeffrey models for viscoelastic solids or liquids). }
    \label{fig:framework}
\end{figure*}

Researchers have come up with a myriad of techniques to navigate this law from images alone~\cite{polacheck_measuring_2016, roca-cusachs_quantifying_2017}. 
Traction Force Microscopy (TFM, Figure~\ref{fig:framework}) is perhaps the most widespread~\cite{legant_measurement_2010}. In TFM, a spatial map of the traction forces exerted by a cell is estimated from the deformation of its substrate, which is first imaged at rest~\cite{han_traction_2015,steinwachs_three-dimensional_2016}. %
Other proxies to measure forces are the deformation of well-characterized droplets embedded in tissue~\cite{campas_quantifying_2014}, %
the orientation or motion of microtubule-kinesin mixtures in active-nematics (AN) systems~\cite{ellis_curvature-induced_2018,boquet-pujadas_inverse_2023}, and the fluorescence of calibrated molecular probes~\cite{blakely_dna-based_2014}. In a reversal of roles, the deformation of biological materials imaged under known conditions of force or stress can help characterize their properties; for instance, through magnetic droplets~\cite{serwane_vivo_2017}, Brillouin microscopy~\cite{scarcelli_noncontact_2015, prevedel_brillouin_2019}, or free deformations of the nucleus~\cite{ghosh_image-based_2021,kesenci_estimation_2024}. Other physical quantities that have also been estimated from images are 
intracellular pressure~\cite{gomez-martinez_silicon_2013, boquet-pujadas_bioflow_2017}, cytoplasmic streaming~\cite{mittasch_non-invasive_2018, klughammer_cytoplasmic_2018}, and tissue stress in monolayers or organs on a chip~\cite{tambe_collective_2011, boquet-pujadas_4d_2022}. From here on, we will talk about ``forces'' for concreteness, but everything applies generally to any of these measurements.

All these methods have led to important discoveries~\cite{kim_embryonic_2021,seelbinder_nuclear_2021,lv_active_2024,head_spontaneous_2024,pallares_stiffness-dependent_2023}, but---besides image data---they have something else in common: They do not report measurement errors despite this being a crucial duty of experimental science. 
The complexity behind the measurement techniques is at the root of this problem. Specifically, most of them involve a first step whereby the deformation $\v(\x)$ is estimated from a pair of images with brightness values $\im_1(\x)$, $\im_2(\x)$ before being fed to a second algorithm (Figure~\ref{fig:framework}, red arrows).
This second step then outputs the spatial map of forces $\f(\x)$ by inverting a physical model defined through some equation system $m(\v; \f)=0$ that relates $\v(\x)$ to $\f(\x)$~\cite{holenstein_high-resolution_2017}. %
Such two-step approaches complicate tracking the propagation of errors and can even amplify them.

To palliate the absence of measurement errors, researchers rely on  
replicates~\cite{vaux_replicates_2012}. Usually, a single figure of merit, such as the mean magnitude $\hat{f}_k = \mathrm{mean}_\x \{ |\f_k(\x)| \}$ of the force, is derived from the spatial map of each cell $k$ and then averaged over a number $n$ of cells or experiments, $\sum_{k=1}^n \hat{f}_k/n$. However, this captures the variability across cells instead of the uncertainty of the measurements. As illustrated by the classical dichotomy between accuracy and precision~\cite{waters_accuracy_2009}, measurements can be consistently wrong. 
Another disadvantage of this approach is that it discards spatial information. As a result, it is difficult to assess the reliability of any pattern that arises in a force map, especially within a single experiment. Without a quantification of its uncertainty, the evidence is circumstantial and vulnerable to computation artefacts~\cite{antun_instabilities_2020}.  %

The aims of this article are twofold. (1) To assign errors to this kind of measurements, and, more importantly, (2) to leverage the resulting credible regions to 
assess the statistical significance (think p-values) of concrete biophysical questions. For example: Did the force pattern change over time? What about after adding a drug? Is the force patch in the measurement map an artefact of the image noise, or is it an actual biological structure of the cell that could correlate with a protein of interest? In this way, researchers will be able to inquire whether what they are observing is statistically significant. 

We provide a framework that achieves (1)-(2) while being general to many mechanobiology measurements (such as to those listed above). 
To illustrate this concisely yet broadly, we have relied on experiments from two measuring techniques---TFM and the imaging of AN systems---with 
disparate 
underlying models: a solid and a fluid, respectively.

\section{Measurement Error}

We consider three main sources of error---image noise, ill-posedness, and model mismatches---and ask how these propagate through the algorithms. We characterize this through the covariance and through credible regions.

\subsection*{A General Formulation of Image-Based Mechanobiology}

To this end, we first argue that most measurement techniques in the introduction can be (alternatively) regarded as stand-alone inverse problems (Figure~\ref{fig:framework}, blue arrows). Contrary to the current two-step approaches, this allows a direct connection from image brightness to force measurements. In~\cite{boquet-pujadas_reformulating_2022}, we developed this idea for TFM. More~generally, we can write our measurement as the $\map(\x)$ solution of an inverse problem constrained by the corresponding physical model $m(\v; \f)=0$ in the form of a PDE system:
\begin{equation}\label{eq:inverse}
 \map = \arg  \min_{\f \in \fspace} \, \big( \of\{\v; \im_{1,2}\} + \reg\{\f\} \big) \text{ subject to } m(\v; \f)=0,
\end{equation}
where $\v(\x)$ and $\f(\x)$ interact through $m=0$. 
Here, the minimization of the optical-flow image-data term
\begin{equation}
\of\{\v; \im_{1,2}\} \approx \sigma^{-2}\int_{\fov} \left(\im_2(\x)-\im_1(\x-\v(\x))\right)^2  \mathrm{d}\x
\end{equation}
favors deformations $\v(\x)$ that best explain the movement that occurred between the two images, $\im_1(\x)$ and $\im_2(\x)$, in the presence of Gaussian noise with a standard deviation $\sigma$ (see Appendix~\ref{sec:inverse_explanation}). On the other hand, the \textit{regularization} term $\reg$ incorporates prior information about the solution (e.g., continuity or smoothness) 
to limit the number of possible solutions to the problem 
(ill-posedness); e.g., $\reg \propto \int_{\fov}\f^2$. 
Therefore, the optimization problem \eqref{eq:inverse} seeks the force map $\map(\x)$ among all possible maps in $\fspace$ that best matches the image data $\im_{1,2}(\x)$ (via $\of$) while satisfying the physical model (via $m$) and maintaining a certain degree of \textit{regularity} (via $\reg$) over the field of view ($\x \in \text{FOV}$). Find examples of these terms in Appendix~\ref{sec:inverse_explanation}. %

We propose that many of the measurement methods listed in the introduction can be expressed in the form of \eqref{eq:inverse} by choosing the corresponding PDE model $m=0$ to match either: an elastic solid for the substrate or the nucleus in TFM or elastography~\cite{han_traction_2015,ghosh_image-based_2021, boquet-pujadas_multiple_2018, kesenci_estimation_2024}, a Stokes fluid for the microtubule-kinesin suspensions or the cytoplasm~\cite{klughammer_cytoplasmic_2018,boquet-pujadas_inverse_2023}, a viscoelastic solid for tissue~\cite{boquet-pujadas_4d_2022}, or a surface tension model for oil droplets~\cite{campas_quantifying_2014}. See Figure~\ref{fig:framework}.

\begin{figure*}
    \centering
\includegraphics[width=\linewidth]{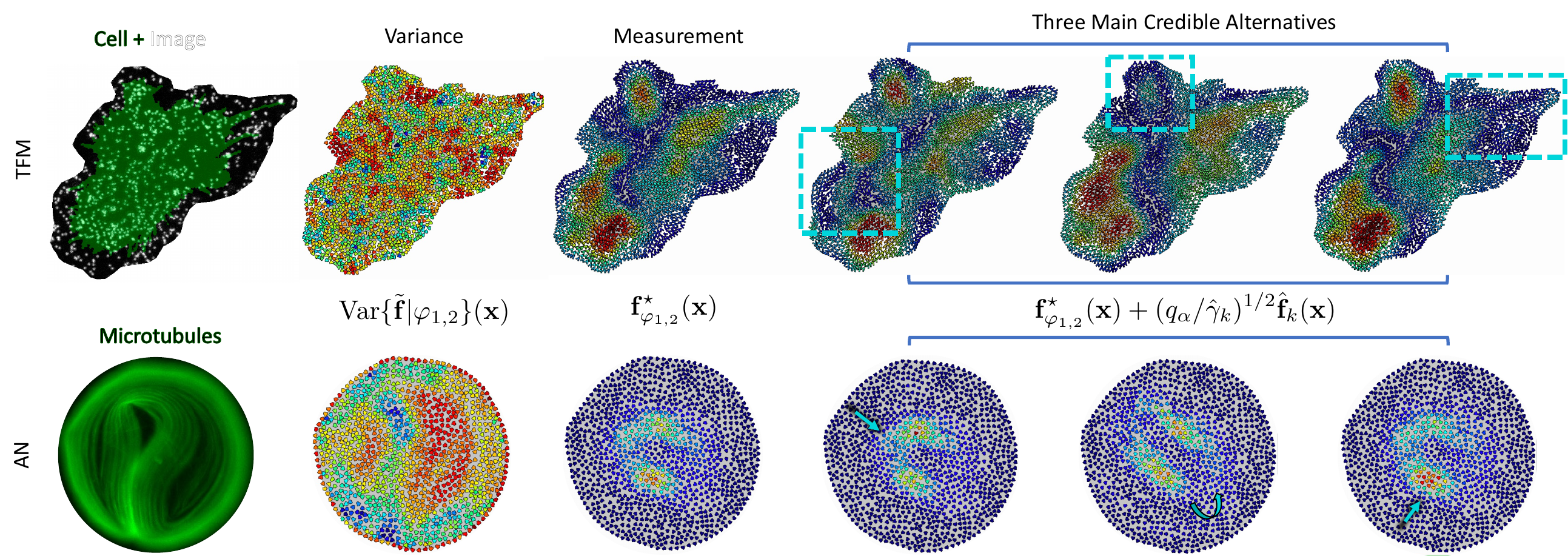}
    \caption{\textbf{Visualization of the uncertainty of the measurements via the variance and MCAs.} \textbf{Top row:} TFM. Left to right: fluorescence image of the beads in the substrate (grey scale) with the cell overlaid (green); variance of the measurement to be compared with the distribution of beads; measurement; and three Main Credible Alternatives (MCAs) to the measurement. Colorbar ranges: \unit{0-8.2\cdot 10^{-4}}{\micro \meter^{-2}} and \unit{0-0.16}{\micro \meter^{-1}}. \textbf{Bottom row:} same as Top but for an AN system. The colorbar ranges are \unit{0-1.5\cdot 10^{-4}}{\micro \meter^{-2} . \second^{-2}} and \unit{0-5.6\cdot 10^{-2}}{\micro \meter^{-1} . \second^{-1}}. Cyan arrows point at relevant differences with respect to the reconstructed measurements.}
    \label{fig:variance}
\end{figure*}

Two advantages of our reformulation over a two-step approach (cf. blue and red arrows in Figure~\ref{fig:framework}) are a reduction in the propagation of errors, and the need for prior information about only one variable---the one that is physically motivated, no less---because the model already constrains which deformations are possible~\cite{boquet-pujadas_reformulating_2022}. 
Since~\eqref{eq:inverse} turns intensity directly into measurements, a third (and most important) advantage is that it can also turn image noise into measurement error.

\subsection*{The Covariance Provides Vectorial Measurement-Error `Bars'}
More precisely, we propose computing the covariance and credible regions of the force from a Bayesian perspective~\cite{robert_bayesian_2007}. For most cases in mechanobiology, where the noise is Gaussian and the model linear (see Appendix \ref{app:nonlinear} otherwise), we can directly work out the covariance $\covt\{\frandc\}(\x)$ through its relation with the Hessian of the functional of the inverse problem in~\eqref{eq:inverse}. Considering the potential measurement force maps $\f(\x)$ as realizations of a random variable $\frandc$ that stems from noise in the images (see Appendix~\ref{sec:inverse_explanation}), we write
\begin{equation}\label{eq:covariance}
    \covt\{\frandc\}(\x) = \hess^{-1}_{\im_{1,2}}(\x),
\end{equation}
\begin{equation}\label{eq:hessian}
    \hess_{\im_{1,2}}(\x) = \hesst \{\of +  \reg \}(\map)(\x),
\end{equation}
where $\hesst \{ \cdot\}$ is the Hessian operator taken with respect to $\f$ (and thus in consideration of $m$), and $|$ conditions the probabilities to the fact that we did acquire the two images.%

The variance $\vart\{\frandc\}(\x)$---which is %
part of 
the covariance 
$\covt\{\frandc\}(\x)$---is a vector field that assigns a (vectorial) standard deviation to the reconstructed force map $\map(\x)$, which is the force map~\eqref{eq:inverse} with the highest probability of explaining the acquired data. %
This means that, unlike in a standard TFM algorithm, the force map is assigned a vectorial error bar at each spatial point $\x$ of the FOV (Figure~\ref{fig:variance}). %

Moreover, the final measurement is no longer a single force map $\map(\x)$, but rather a full probability distribution $\den_{\frandc}(\f)$ that assigns a probability to every possible force map $\f(\x) \in \fspace$ given the fact that we did acquire two specific images $\im_{1,2}$. An expression for $\den_{\frandc}(\f)$ can be found in Appendix~\ref{sec:credibleregion}. Hereafter, we drop the spatial dependency of functions on $\x$ for conciseness.

\subsection*{Credible Regions}
To better visualize the uncertainty of the measurements and to set up our hypothesis-testing framework (Section~\ref{sec:hypothesis-testing}), we now turn to the so-called credible regions of the probability density $\den_{\frandc}(\f)$.

To define them, we seek the force maps with the highest probabilities of explaining our acquisition of images $\im_1$ and $\im_2$. %
More specifically, we consider the smallest set $\ci^\alpha_{\frandc} \subset \fspace$ of force maps $\f$ such that their probabilities together sum to $1-\alpha$, i.e.
\begin{equation}\label{eq:prob}
    \pro_{\frandc} \left[ \f \in \ci^\alpha_{\frandc} \right] \geq 1 - \alpha,
\end{equation}
where $\alpha$ is our pre-established confidence level. 
The information to build this credible set (or region, or interval) $\ci^\alpha_{\frandc}$ is contained in our density $\den_{\frandc}(\f)$. (Find an example in Appendix~\ref{sec:credibleregion}.) By definition, the most probable force map $\map$ given the image data and ill-posedness is in this set of the most probable force maps. 

Relative to the significance level $\alpha$, the credible regions capture the variability of all the potential measurements that could be behind the (noisy) acquired images $\im_{1,2}$. In Appendix~\ref{sec:credibleregion}, we provide an expression for $\ci^\alpha_{\frandc}$ and show how to leverage the eigenvectors of $\hess_{\im_{1,2}}$ to visualize it. The visualization consists of a few alternative force-measurement maps \eqref{eq:eigen_regions} 
the distribution of which has been altered by changing the characteristics that we are most uncertain of (up to some level $\alpha$). By comparison to $\map$, they highlight what aspects of our measurement we are least certain about. We will refer to these maps as MCAs for Main Credible Alternatives (Figure~\ref{fig:variance}). Next, we show how this $\alpha$ relates to p-values.

\section{Hypothesis Testing}\label{sec:hypothesis-testing}
Besides providing the variance or the standard deviation, 
the probability distribution $\den_{\frandc}(\f)$ of the measurement opens the floor to statistical tests; for example, through the credible regions in~\eqref{eq:prob}. Imagine that we have an hypothesis about some biological forces (see example questions below); and that we then perform the necessary experiments, acquire our measurements in the form of images, and reconstruct the forces therefrom. How likely is our hypothesis to be ``correct'' given the outcome of the %
image acquistion 
when considering the noise and the ill-posedness of the reconstruction? 
Here, we present a framework to test hypotheses in such a context. 

Much like in classical testing, the overarching paradigm from here on will be to `favor' our proposed (alternative) hypothesis $H_1$ by rejecting the opposite (null) hypothesis $H_0$ with a level $\alpha$ (p-value) of pre-established confidence~\cite{casella_statistical_2002}.

\subsection*{A Framework for Testing}
Based on the credible regions in~\eqref{eq:prob}, we address the statistical significance of experimental questions by examining the overlap 
\begin{equation}\label{eq:intersection}
     \null \cap \ci^\alpha_{ \frandc} 
\end{equation}
between the credible set and the set of force maps $\null \subset \fspace$ that are associated to the null hypothesis $H_0$ under investigation. (For example, $\null = \{ \mathbf{0}\}$ when $H_0$ posits that no force is present.) 
If the sets do not intersect, we reject the null hypothesis with an
$\alpha$ that can be interpreted as a p-value in the sense that we demonstrate in Appendix~\ref{sec:interpretationalpha}. We remark that we have chosen our test~\eqref{eq:intersection} to err on the side of caution, favoring false negatives (type II errors) over false positives (type I errors).

\subsection*{Testing of Research Questions}
We now present several example questions.

\begin{figure}
    \centering
    \includegraphics[width=\linewidth]{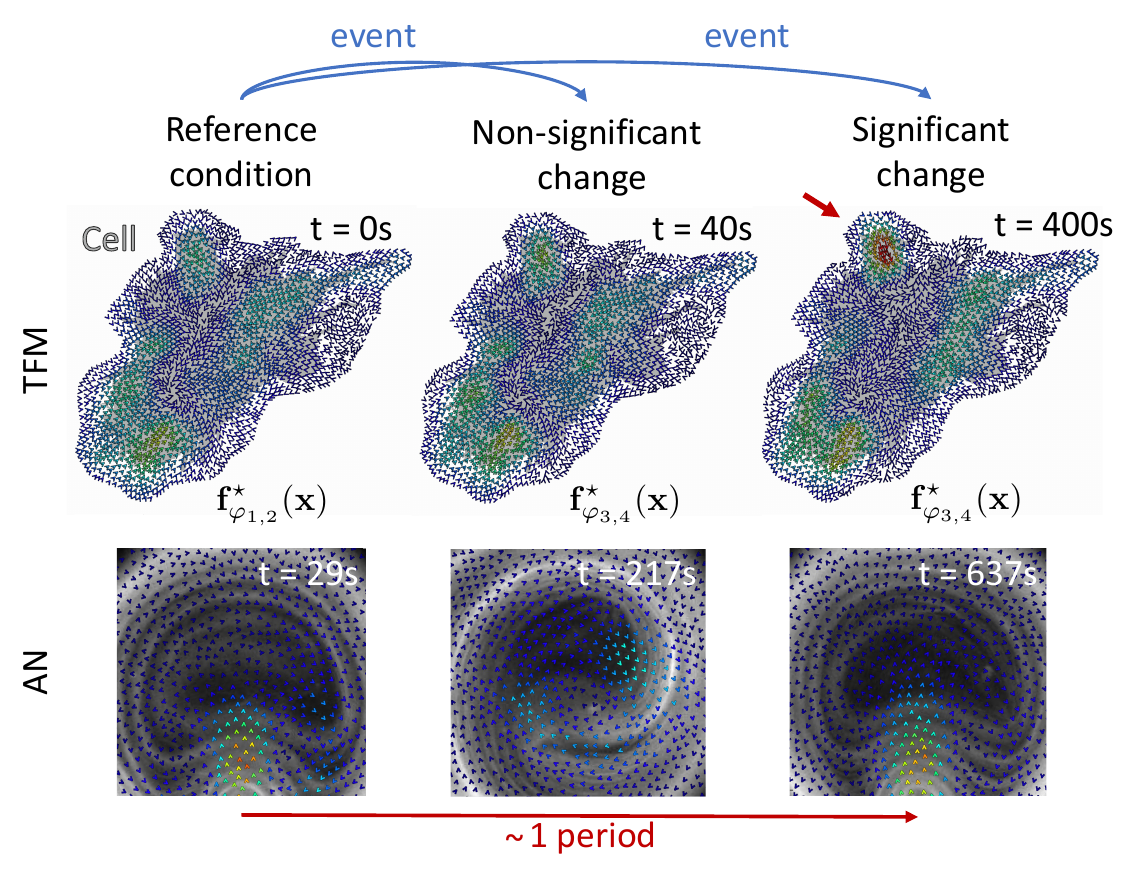}
    \vspace{-20pt}
    \caption{\textbf{Hypothesis tests for the significance of force changes after some event.} (Question 1.) The tests show that some events are not significant enough with respect to the ill-posedness and image noise. \textbf{Top row:} TFM. Colorbar ranges are \unit{0-0.16}{\micro \meter^{-1}}. Left to right: force maps at {0}{\second}, {40}{\second} (non-significant change), and \unit{400}{\second} (significant change) as the cell establishes after seeding. \textbf{Bottom row:} same as Top but for an AN system. The colorbar range is \unit{0-2.8\cdot 10^{-2}}{\micro \meter^{-1} . \second^{-1}}.}
    \label{fig:differences}
\end{figure}

\medskip
\textit{Question 1:} We would like to first ask whether the force map in a cell changes significantly after some event, for example in response to a drug or, more simply, as time goes by (Figure~\ref{fig:differences}). To address this question, we take a measurement $\im_{1,2}$ before the event and another one, $\imm$, after.

Note that each measurement consists of two images. For each pair, we can use \eqref{eq:inverse} to compute their respective highest-probability force maps, $\map$, $\f^\star_{\imm}$; and use \eqref{eq:covariance}-\eqref{eq:hessian} to ``obtain'' their densities, $\den_{\frand|\im_{1,2}}(\f)$, $\den_{\frand|\imm}(\f)$. 

Our expectation, and therefore our alternative hypothesis $H_1$, is that the force pattern changed after the event. Accordingly, the formulation of our null hypothesis is that the spatial distribution of the force did not change, i.e. $H_0: \map = \f^\star_{\imm}$.

By~\eqref{eq:intersection}, we can then consider that the event changed the force pattern significantly ($\alpha)$ when
\begin{equation}\label{eq:q1}
 \{ \mathbf{0} \}  \cap \ci^\alpha_{\frandc-\frand |\imm} = \emptyset,
\end{equation}
or, equivalently, when
\begin{equation}\label{eq:q1}
\mathbf{0} \notin \ci^\alpha_{\frandc-\frand |\imm}.
\end{equation}
This can be determined by performing a computation with the Hessian %
and the corresponding chi-squared quantiles, as described in Appendix~\ref{sec:tests}. There, we also show how to restrict the test to a small zone of interest (e.g., a single adhesion patch) within the cell and provide a connection with the classical Wald~test.

\medskip
\textit{Question 2:} %
One could also question whether the force recovered from a pair of images $\im_{1,2}(\x)$ is significant at all. In other words, is the force distinguishable from the background? 
To answer this, one can check whether
\begin{equation}\label{eq:background}
 \{ \vb^\star \}  \cap \ci^\alpha_{\frandc-\brand} = \emptyset,
\end{equation}
where $\vb^\star(\x)$ represents the expected background and $\brand$ corresponds to a noise model with zero mean. (see Appendix~\ref{sec:tests} for the necessary computations.) If~\eqref{eq:background} holds, we reject the hypothesis that the force looks like the background with a confidence level~$\alpha$.

\begin{figure*}
    \centering
    \includegraphics[width=\linewidth]{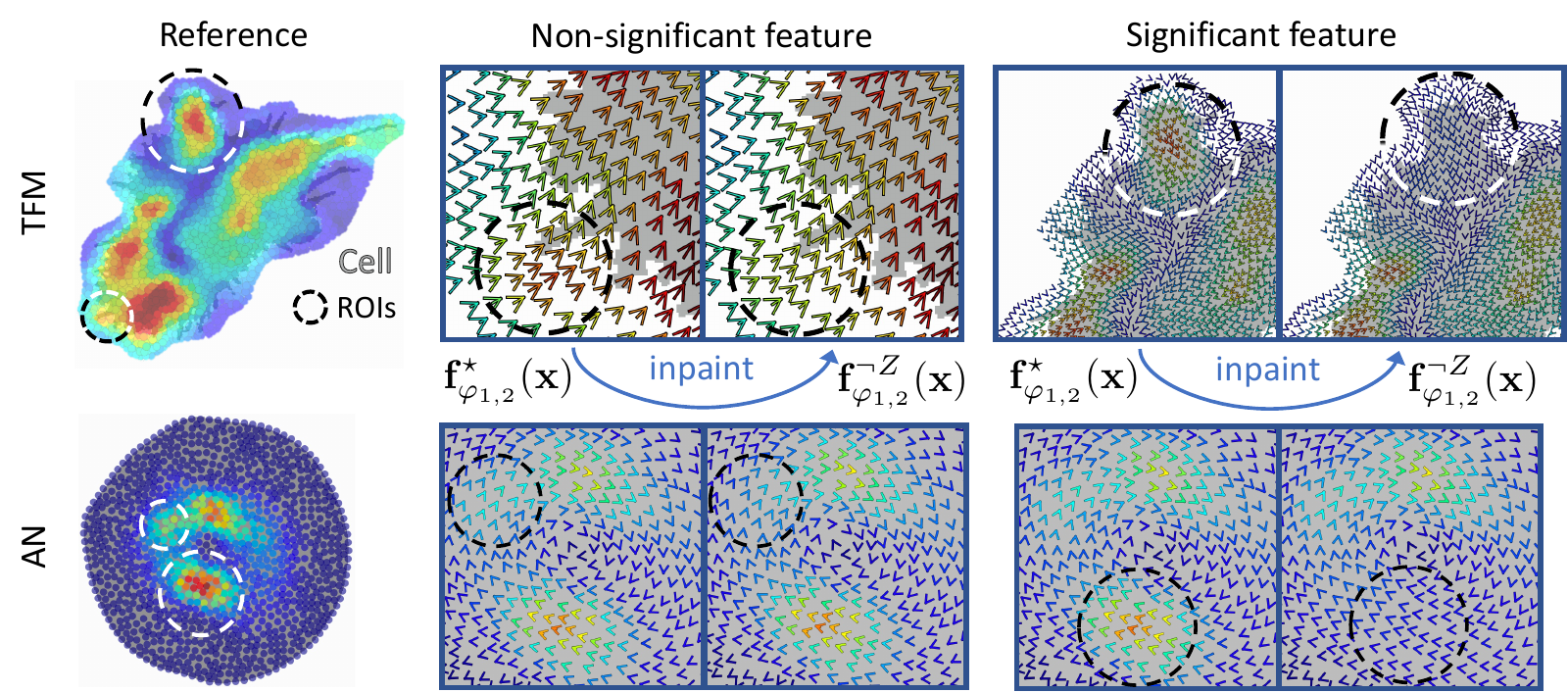}
    \caption{\textbf{Hypothesis tests for the significance of features in the measurements.} (Question 3.) The tests show that the existence of some force patches is uncertain when noise and ill-posedness are accounted for. \textbf{Top row:} TFM (range \unit{0-0.16}{\micro \meter^{-1}}). Left to right: A map of the magnitude $||\map||_2(\x)$ of the measurement overlaid on the cell with circles delimiting two regions of interest (ROIs) for reference. Zoom-ins around one of the ROIs showing a comparison between the original reconstructed force field and a version where we inpainted the ROI (change not significant). Zoom-ins around the other ROI for which the inpainting did result in a significant change. \textbf{Bottom row:} same as Top but for an AN system (range \unit{0-5.6\cdot 10^{-2}}{\micro \meter^{-1} \second^{-1}}).}
    \label{fig:inpainting}
\end{figure*}

\medskip
\textit{Question 3:} 
We may not always be able to model the background, especially when it is not uniform. For example, imagine that we have acquired $\im_{1,2}$ and wish to assess whether the presence of a small force patch inside a cell is statistically significant, or if it could simply be a product of noise or randomness (Figure~\ref{fig:inpainting}).

In this scenario, our alternative hypothesis is that a force patch is present, whereas the null hypothesis posits that there is none. 

To conduct this test in the absence of a background model, we propose generating an alternative measurement map without the force patch. We do this formally by segmenting and interpolating (inpainting) the region of interest $Z \subset \text{FOV}$ (see Appendix~\ref{sec:inpainting}). The result is a set $\inp$ of alternative force maps that do not include the force patch. The null hypothesis is then that the measurements resemble those in $\inp$, and we reject it in favor of the alternative hypothesis with significance level $\alpha$ when
\begin{equation}\label{eq:test_inpaint}
\inp \cap \ci^\alpha_{\frandc} = \emptyset.
\end{equation}
The computations necessary for this test, along with an additional example question, can be found in Appendix~\ref{sec:tests}.

\medskip
\textit{Significance in practice:} Hypothesis tests are designed to assess \textit{statistical significance}. However, their interpretation in practical terms has long been a subject of debate~\cite{shi_reconnecting_2021, chen_roles_2023}. In Appendix~\ref{sec:practical}, we discuss how to also incorporate \textit{practical significance} to our statistical tests. Our proposal is to replace the single-point null hypotheses $\{0\}$, $\{\vb^\star\}$ in questions 1 and 2 with (ball) sets of ``close-enough'' force-map alternatives. We find that this approach effectively intertwines both significances in a constructive way similar to equivalence-test procedures.%

Details about the implementation of these methods, including the algorithm, can be found in Appendix~\ref{sec:implementation}. %

\section{Experiments}

\paragraph*{General Framework} In Figure~\ref{fig:framework}, we present examples of the measurement techniques (TFM, AN) used to illustrate our general framework. Appendix~\ref{sec:experiment_details} provides the technical details of these experiments.

In the TFM example, a cell exerts traction forces on a polyacrylamide substrate~\cite{sune-aunon_full_2017,jorge-penas_free_2015}, which behaves as an elastic solid (Figure~\ref{fig:framework}, top). The data consist of two images: one where the substrate is at rest and the other where it is deformed. Using our framework (Figure~\ref{fig:framework}, blue arrows), we can reconstruct the displacements and the force at once. This is in contrast to the classic two-step approach (red arrows), wherein one first needs an algorithm to compute the displacements before using an \textit{ad hoc} method to invert the physical model. We documented the advantages of our approach in terms of the accuracy of TFM in~\cite{boquet-pujadas_reformulating_2022}. In the present instance, the forces that we measure from the image data using \eqref{eq:inverse} show that the traction is localized at certain protrusions of the cell as it pushes inward (Figure~\ref{fig:framework}, top, rightmost).

Our framework can reformulate other measuring techniques that also rely on first measuring ``deformations'' (displacements, velocities, changes in curvature, etc.) from images (Figure~\ref{fig:framework}, bottom). Taking measurements in AN systems provides another example (Figure~\ref{fig:framework}, second row). In this work, we consider microtubulin-kinesin mixtures, which are driven by the forces generated by these ATP-powered molecular motors. The resulting dynamical systems are forced Stokes fluids. These ones in particular behave as follows: Two $+1/2$ defects rotate around the center, driving a system-wide circular flow. This motion is periodically and temporarily disrupted by a $+1/2$ defect, which is propelled toward the center after nucleating at the boundary. This defect eventually replaces one of the rotating defects, which merges with the $-1/2$ left behind at the boundary. The data we use for measurements consist of two consecutive images of the system~\cite{opathalage_self-organized_2019}. In this case, we use a Stokes model for $m=0$, with velocity playing the role of displacements in TFM. Our measurements capture the system's velocity everywhere, including at the boundaries, as well as the forces involved, which result from the divergence in the alignment of microtubules and lead to bending instabilities. For example, the measurements in Figure~\ref{fig:framework} are taken just before the defects begin their circular motion.

\paragraph*{Measurement Error} Through the direct link between image data and physical measurements, our framework quantifies measurement error.

Importantly, in simulated TFM experiments, where a perfect ground truth was available for comparison, we found that the variance computed via~\eqref{eq:covariance} reflected the error between the recovered force $\map(\x)$ and the true force~\cite{boquet-pujadas_reformulating_2022}. 

Moreover, our observations suggest that regions with less contrast or texture lead to more uncertain measurements, as do images corrupted by higher noise. This is true for both TFM simulations~\cite{boquet-pujadas_reformulating_2022} and real data from TFM and AN systems (here). For example, zones without fluorescent beads in the TFM data (Figure~\ref{fig:variance}, top row, first two columns) show higher variances due to a lack of information. Similarly, regions with weaker brightness gradients exhibit greater uncertainty in the AN system (bottom row, first two columns), though this also interacts with other important physical sources of information, such as the incompressibility of the fluid. The variance also highlights the most informative regions, effectively serving as a sensitivity analysis for experimental design. For instance, one could adjust the density of fluorescent beads in TFM to optimize accuracy.

\def\pv{10^{-3}}
\paragraph*{Credible Regions and Alternatives} The uncertainty of the measurements is difficult to visualize using the variance alone, especially since the full covariance must be considered. To address this, we leverage the MCA force maps in~\eqref{eq:eigen_regions} to explore the credible region of the posterior. We show three of these probable ($\alpha=\pv$) alternative maps in Figure~\ref{fig:variance}.

In the case of TFM, these alternatives show how much the force patches can vary and how confident we can be about their existence. In particular, we can be more confident that the top force patch exists (it never fades out at this $\alpha$ level, even for the very specific MCA in the middle), but less so about the other two patches (left and right MCAs). In short, these MCAs tell us the plausible ways in which the force map can vary without contradicting the ill-posedness or the noise range of the image acquisition.

In the case of the AN system, the MCAs show that we can be fairly certain there are two force spots pointing in opposite directions. However, some aspects of the measurement remain uncertain. For example: the presence of a small force tail on the top force patch (Figure~\ref{fig:variance}, leftmost MCA); the overall direction of the two opposing patches (the middle MCA is rotated but remains a potential alternative); or the possibility that one of the opposing patches may have a greater magnitude than the other (left and right MCAs). Therefore, experimental conclusions based on any of these uncertain features may not be well-supported.

To formalize the kind of questions that follow organically from this MCA analysis, we rely on our hypothesis-testing framework~\eqref{eq:intersection}.

\paragraph*{Hypothesis Testing Q1}  In Figure~\ref{fig:differences}, we illustrate Question 1; i.e., we ask whether the measurements change significantly after some event. We set the significance level necessary to reject the null hypothesis to $\alpha=\pv$.

For the TFM example, we compare the traction forces of a wild-type cell measured after being seeded ($t=\unit{0}{\second}$) with those measured after the cell starts to establish itself at $t=\unit{40}{\second}$, and $t=\unit{400}{\second}$. The resulting force maps indicate that the cell appears to pull more strongly from the top force patch as time goes by (Figure~\ref{fig:differences}, top, red arrow). Running test~\ref{eq:q1} reveals that the difference is only significant at the later time point of $t=\unit{400}{\second}$. By contrast, the smaller increase in traction at $\unit{40}{\second}$ is not $\alpha$-significant at this level of noise and ill-posedness. 

For the AN system, applying test~\eqref{eq:q1} to compare two force measurements taken at different phases of the system's dynamics yielded a significant difference. Conversely, comparing two (carefully selected) measurements taken during the same phase but at different periods showed no significant difference (Figure~\ref{fig:differences}, bottom).

\paragraph*{Hypothesis Testing Q2} All the force maps in Figure~\ref{fig:differences} tested significant relative to the background as per the test~\eqref{eq:background} formulated for Question 2.

\paragraph*{Hypothesis Testing, Q3}

Finally, in Figure~\ref{fig:inpainting}, we explore whether certain features of the measurement itself (without reference to any event) are significant. This corresponds to Question 3. 

In the TFM example, we focus on two notable hotspots of traction force that coincide with different protrusions of the cell (Figure~\ref{fig:inpainting}, top). We do not find the smaller hotspot significant upon application of our test~\eqref{eq:test_inpaint}. However, the larger hotspot at the top of the cell does pass the test. This suggests that basing any conclusion on the existence of the smaller hotspot (say a correlation with a protein or with the protrusion itself) would be less sensible, given the ill-posedness and the noise in the image data. By contrast, we can be more confident in the existence of the larger force patch. These conclusions are consistent with the observations made using the MCA visualization: In Figure~\ref{fig:variance}, the smaller hotspot was much more keen to disappear (left MCA), whereas the larger one could only be partly weakened using a very specific direction $\hat{\mathbf{f}}_k$ (middle MCA). 

For the AN example (Figure~\ref{fig:inpainting}, bottom), the two features that we tested were a small tail preceding the top patch, which we found insignificant, and the entire bottom patch, which was significant.

\section{Discussion}
We have presented a framework that allows associating a significance level to common experimental questions. A key step was rewriting several methods in the field of mechanobiology under a common formulation based on a one-step inverse problem that is fed directly by images. This allowed us to study how image noise propagates into measurement errors. The visualization of the resulting probability space through the Main Credible Alternatives highlighted force features that were uncertain and prompted further research questions. To formalize the asking of these questions, we developed a framework for the testing of biophysical hypotheses.

Our observations with hypothesis testing suggest that caution is needed in the interpretation of measurements taken with image-based techniques, especially when the structures of interest are small or coincide with zones of low image contrast. In our experiments, some such structures did not pass the significance test for several reasons. One factor is noise, which contaminates the acquisitions. Another is the ill-posedness of the reconstruction problem itself, whereby many deformations can cause similar changes in the images because information is only available in the direction of the image gradient. Therefore, we recommend putting more effort into accounting for these factors when taking measurements in biophysical experiments, which often involve (two-step) inverse problems due to the inaccessibility of the sample. The overarching goal of this work has been to augment the communication of experimental conclusions by associating a confidence level thereto. %
We believe that the specific p-value and effect-size thresholds should be established through a community-wide discussion.

An alternative application of the ideas behind our work could involve evaluating the performance of computational imaging systems with respect to the subsequent inverse problem. This evaluation would be integrated into the simulation stage of the design process, complementing best-case scenario metrics such as lower Cramér-Rao bounds. For example, it could help answer questions like whether a super-resolution microscope will be able to resolve a specific compound of interest~\cite{schermelleh_super-resolution_2019,hockmann_analysis_2024}, or whether a prototype PET scanner will be able to detect a small atherosclerotic plaque~\cite{doyon_first_2023, boquet-pujadas_silicon-pixel_2024}. A key objective in these applications would be to achieve statistical significance while minimizing cell phototoxicity or radiation dose.

Other future directions include designing experiments that minimize uncertainty, developing more general Monte-Carlo-based algorithms (see Appendix~\ref{app:nonlinear}), exploring applications in other areas of physics such as calcium or black-hole imaging~\cite{zhang_fast_2023,pham_deep-prior_2024,the_event_horizon_telescope_collaboration_first_2019}, refining the tests to better tune the balance between type I and type II errors, and integrating practical and statistical significance more~seamlessly.

\section*{Acknowledgments}
The author is grateful to Sangyoon J. Han and Nikhil Mittal, as well as Arrate Mu\~noz-Barrutia and Miguel Vicente-Manzanares, for sharing TFM raw data. This work was funded (in part) by the Swiss National Science Foundation under the Sinergia grant CRSII5 198569.

\appendix%
While we keep the same notation as in the main text for clarity, throughout the Appendix we interpret the variables as discrete vectors rather than as functions for simplicity. (Note that many of the quantities, such as force or displacements, are originally vector fields, which we discretize and then vectorize into discrete vectors.) A detailed explanation of this notation can be found in Appendix~\ref{sec:notation}. The implementation details for the concepts discussed in this appendix are provided in Appendices~\ref{sec:implementation} and~\ref{sec:experiment_details}.

\subsection{The Inverse Problem}\label{sec:inverse_explanation}
\subsubsection{Deterministic Data Term}
\label{sec:of}
The optical-flow term in~\eqref{eq:inverse} incorporates image data into the inverse problem. In particular, it has the role of estimating a motion or deformation map $\v(\x)$ from two consecutive images by assuming that the brightness (e.g., from protein emissions) is conserved but redistributed within the image, i.e. $\im_2(\x)-\im_1(\x-\v) \approx 0$. One can write this variationally as
\begin{equation}
    \hat{\of}\{\v; \im_{1,2}\} \approx 0
\end{equation}
with
\begin{align}\label{eq:of_nonconvex}
&\hat{\of}\{\v; \im_{1,2}\} = ||  \im_2(\x)-\im_1(\x-\v)||^2_{\fov, 2, \mathbf{W}} \\ 
&=   \int_{\fov} (  \im_2(\x)-\im_1(\x-\v)) \mathbf{W} (\im_2(\x)-\im_1(\x-\v)) \, \mathrm{d}\x. \nonumber 
\end{align}
Here, $\mathbf{W}$ is a weighting matrix for the inner product (or for the norm). This is most often chosen as proportional to the identity, $\propto \mathbf{I}$, and it is meant to balance the effect of the data term with respect to the regularization term.

In most cases, the optical-flow term \eqref{eq:of_nonconvex} is linearized as
\begin{equation}\label{eq:of_convex}
\of\{\v; \inte\}=||\inte+\mathbf{O}\v||^2_{\fov, 2, \mathbf{W}} \approx \hat{\of}\{\v; \im_{1,2}\} 
\end{equation}
to make the problem convex and, thus, easier to solve. In~\eqref{eq:of_convex}, $\inte=\im_2-\im_1$ and $\mathbf{O}$ stands for the inner product with the gradient, i.e. the dot product with $\nabla \im_2$. The linearization \eqref{eq:of_convex} is an approximation of~\eqref{eq:of_nonconvex} that gets more accurate as deformations get smaller. Small deformations are the norm in mechanobiology. One can also make deformations small on purpose by increasing the frame rate of the acquistion, or by decreasing the expected resolution. Therefore, \eqref{eq:of_convex} constitutes a good approximation. If eventually necessary, one can also process big deformations by embedding the framework into a multiresolution scheme whereby \eqref{eq:inverse} is solved over a pyramid of spatial scales~\cite{papenberg_highly_2006,boquet-pujadas_reformulating_2022}, wherein deformations become (relatively) increasingly smaller.

Together, the data term, the regularization term $\reg\{\f\}$%
, and the model constraint $m(\v;\f)$ make up the inverse problem in~\ref{eq:inverse}. The result thereof is a force map $\map$ that fulfills the model, agrees with the image data, and is regular enough with respect to the choice of regularization.

\bigskip
\subsubsection{Regularization Term or Prior}
One widespread example of a Gaussian-inducing regularization term is
\begin{equation}\label{eq:reg}
\reg\{\f\}=\beta ||\mathbf{L}\f||_{\fov, 2}^2,
\end{equation}
where $\mathbf{L}$ is the discretization of some linear operator such as the identity or the gradient, and $\beta \in \reals_{> 0}$ weights the balance between data fidelity and prior information. For our examples we use use~\eqref{eq:reg} with the identity because in TFM and AN it acts akin to a low-pass filter due to the Laplacian-like operator in the PDEs~\cite{boquet-pujadas_reformulating_2022,boquet-pujadas_inverse_2023}. 

\bigskip
\subsubsection{Physical model}
The physical model links the velocity or displacement $\v$, which can be inferred from the data (via OF), to the force $\f$. Other than providing the measurement of interest, this avoids having to split the problem into two steps, and helps motivate the regularization more physically. One example of a model is $m(\v;\f)=0$ with
\begin{numcases}{m(\v;\f)=}\label{eq:elastic}
	\nabla \cdot \left( - p \mathbf{I} + \mu \left(\nabla \v + \nabla^\intercal \v \right ) \right) + \mathbb{1}_\mathrm{K} \f  & \text{in $\Omega$,} \nonumber \\ 
	\nabla \cdot \v  + p / \lambda & \text{in $\Omega$,} \\
	\v - \mathbf{g} & \text{on $\partial \Omega$}, \nonumber
\end{numcases}
where $\mathbf{g}$ is the value of the displacements at the boundary, $p$ a state variable that acts as an auxiliary pseudo-pressure, and $\lambda$, $\mu$ are the Lamé parameters. Note that the shear modulus $\mu$ is related to the Young's modulus and the Poisson's ratio of the material according to $\mu = E/(2(1+\nu))$ in both 2D and 3D. 
We use this model for TFM, where $\mathrm{K}$ is the spatial domain of the cell, and the model domain spans the entire field of view, i.e., $\Omega = \fov$. 

\bigskip
\subsubsection{Bayesian Data Term}
In this work, we regard~\eqref{eq:of_nonconvex} from a (Bayesian) statistical perspective by considering that the images are noisy $\im_2(\x)-\im_1(\x-\v) \sim \mathcal{N}\left(c,  \mathbf{W}^{-1} \right)$ with covariance $\mathbf{W}^{-1}$ and mean $c$. Here, we intentionally call the covariance $\mathbf{W}^{-1}$ because it can be identified as the weighting in the deterministic interpretation~\eqref{eq:of_nonconvex}. Usually, the noise is spatially independent and uniform, $\mathbf{W}^{-1}=\sigma^2 \mathbf{I}$, so 
\begin{align}%
\of\{\v(\f); \im_{1,2}\} &= ||\inte+\mathbf{O}\v(\f)-c||^2_{\fov, 2, \mathbf{\sigma^{-2} \mathbf{I}}} \\ 
  &\approx \sigma^{-2}\int_{\fov} \left(\im_2(\x)-\im_1(\x-\v(\x))\right)^2 \,  \mathrm{d}\x. \nonumber 
\end{align}
From the Bayesian perspective, we intepret the data terms~\eqref{eq:of_nonconvex} or~\eqref{eq:of_convex} as the (negative) logarithm of a \textit{likelihood}. In particular, of the \textit{likelihood} $\den_{\imrand|\f} \propto \exp\{-\of\{\vrand(\frand); \im_{1,2}\}\}$ of acquiring two specific images given that the underlying deformation is known; or, more precisely, that the underyling force is known if we remember to consider the model constraint $m(\v;\f)$. Notice that the \textit{likelihood} is the opposite of what we want, which is the probability distribution of $\frand|\im_{1,2}$. Both are related through Bayes rule as per $\den_{\frandc} \propto \den_{\imrand|\f} \, \den_{\frand}$. Indeed, under this interpretation, the regularization term $\reg\{\f\}$ in~\eqref{eq:inverse} acts as a \textit{prior} through $\den_{\frand} \propto \exp\{-\reg\{\f\}\}$. The maximum a posteriori (MAP) of $\den_{\frandc}$---i.e. the most probable force map---is precisely the solution $\map$ of the inverse problem~\eqref{eq:inverse}. Moreover, $\den_{\frand}$ also contains information about how probable other force maps are in light of the images acquired and, thus, reflects measurement error. %

The algorithm to solve the optimization problem in~\eqref{eq:inverse} is detailed in Appendix~\ref{sec:implementation}.

\subsection{The Credible Regions}\label{sec:credibleregion}
To illustrate the mathematical concepts in the main text, we present expressions for the posterior random variable $\frandc$; for its (posterior) density function $\den_{\frandc}(\f)$, which associates a probability to all of the possible force maps; and for the smallest $\alpha$-credible region $\ci^\alpha_{\frand|\im_{1,2}}$, which groups together the force maps with the highest probabilities for the purpose of our tests. %
Recall that throughout all the Appendix we interpret the variables as vectors instead of functions for simplicity (see Appendix~\ref{sec:notation}).

For most cases in mechanobiology, where both the noise and the problem are Gaussian, expressions \eqref{eq:covariance}-\eqref{eq:hessian} are exact. Then, the distribution of the resulting posterior random variable follows 
\begin{equation}\label{eq:distributedas_gaussian}
  \frandc  \sim \mathcal{N}\left(\map,  \covt\{\frandc\} \right),
\end{equation}
where we use a tilde to distinguish random variables from realizations thereof, and we define $\frandc := (\frand |(\imrand=\im_{1,2}))$. 
The density is then
\begin{equation}
\den_{\frandc}(\f) \propto \exp{\left\{-\frac{1}{2} \left( \f- \map \right)^\intercal \hess \left( \f- \map \right)\right\}},
\end{equation}
and the (smallest) credible regions are the ellipsoids
\begin{equation}\label{eq:ci_gaussian}
    \ci^\alpha_{\frand|\im_{1,2}} =  \{ \f : \left( \f- \map \right)^\intercal \hess \left( \f- \map \right) \leq q_\alpha \},
\end{equation}
where $\covt\{\frandc\}=\hess^{-1}$, and $q_\alpha$ marks the $\alpha$-quantile. This is the quantile of the chi-squared distribution $\chi^2_{\fdim}$ with $\fdim$ the dimension of the discretized force map. 
Note that the (smallest) credible region~\eqref{eq:ci_gaussian} is a (convex) level set of the probability density function $\den_{\frand|\im_{1,2}}(\f)$; this is a given under the common condition of convexity. Note also that we sometimes drop the Hessian's subscript for conciseness, $\hess=\hess_{\im_{1,2}}$.

See Appendix~\ref{app:nonlinear} for a discussion of \eqref{eq:distributedas_gaussian}-\eqref{eq:ci_gaussian} in cases where some of the assumptions in this section might not hold. %

\subsubsection{Visualization via Eigenvectors}
To visualize the high-dimensional credible regions in~\eqref{eq:ci_gaussian} we rely on the eigenvectors $\hat{\f}_k$ of $\hess_{\im_{1,2}}$, which are force maps themselves. (These are also the eigenvectors of the covariance because it is the inverse.) 
In particular, the force maps 
\begin{equation}\label{eq:eigen_regions}
\map \pm  (q_\alpha/\hat{\gamma}_k)^{1/2} \hat{\f}_k, 
\end{equation}
point along an axis of the ellipsoid and reach the $q_\alpha$-level set, where $\hat{\gamma}_k$ is the corresponding eigenvalue. %
As a result, the force maps \eqref{eq:eigen_regions} are good representations of the statistical variability of the reconstruction and can help highlight a subset of characteristics of the force distribution of which one should remain uncertain about. (These characteristics are precisely the eigenvector directions.) We call the force maps in~\eqref{eq:eigen_regions} ``Main Credible Alternatives'' because we think of the eigenvectors as acting as main credible directions. See Appendix~\ref{sec:implementation} for the implementation.

\subsection{Interpretation of the Confidence (or Significance) Level}\label{sec:interpretationalpha}
We chose to perform our tests by checking whether the $\null$ set, which encompasses all the possible force maps within the null hypothesis, intersects with the confidence region $\ci^\alpha_{\frand|\im_{1,2}}$, which corresponds to a confidence level $\alpha$~\cite{berger_testing_1987,casella_reconciling_1987}. In this section, we explain how $\alpha$ can be interpreted as a confidence or significance level.

\bigskip

\subsubsection{Bayesian Perspective} If the sets do not intersect ($\null \cap \ci^\alpha_{\frandc} = \emptyset$) for a given $\alpha$, the probability associated with the null hypothesis (upon acquiring the images) is smaller than $\alpha$ and we reject $H_0$. This is because
\begin{equation}
    \pro_{\frandc} [\f \in \null] \leq \pro_{\frandc}\big[\f \in {\big(\ci^\alpha_{\frand|\im_{1,2}} \big)}^\mathsf{c}\big]=\alpha
\end{equation}
since $\null \subset {\big(\ci^\alpha_{\frandc} \big)}^\mathsf{c}$. This probability estimate might seem loose, but it is generous towards minimizing type I errors\footnote{This is in line with the ethical choice in law courts (when the null hypothesis implies innocence), with the stricter choice in clinical trials (when the null hypothesis implies a non-effective pharmaceutical drug), and with the safest choice in medicine (when the null hypothesis implies illness).}, which means that our research hypothesis (the alternative one) will only be backed when there is enough convincing evidence. (Notice that to obtain the opposite effect, one can swap the null and alternative hypotheses.) If the sets do intersect, \mbox{$\null \cap \ci^\alpha_{\frandc} \neq \emptyset$}, we cannot reject the null hypothesis.

\bigskip

\subsubsection{``Extreme'' Perspective}
\def\aalpha{\hat{\alpha}}
Instead of prescribing an $\alpha$ a priori, one could look for the smallest $\alpha$ (or the biggest credible region) such that there is no intersection 
\begin{equation}
\aalpha = \min\{\alpha \in (0,1)|\null \cap \ci^\alpha_{\frandc} = \emptyset\}.    
\end{equation}
Then $\aalpha$ is the probability associated to any elements that are as likely or less likely than those that conform the null hypothesis. This reinforces the preference for controlling type I errors, and ties well with the meaning behind the classical frequentist p-value, which is based on the probability of observing elements at least as ``extreme'' as the one that was observed experimentally. For example, with the working credible region $\eqref{eq:ci_gaussian}$, one can find $\aalpha$ by first computing the quantile 
\begin{equation}
q_{\aalpha}=\min_{\f \in \null}\left( \f- \map \right)^\intercal \hess \left( \f- \map \right)
\end{equation}
of the chi-squared distribution and then finding the associated probability using the cumulative function.

\bigskip

\subsubsection{Remark} Notice that the set of force maps associated to the alternative hypothesis $H_1$ is the complement with respect to $\fspace$ of the set~$N$ associated with the null hypothesis $H_0$. In other words, it contains all the possible force maps that are not in~$N$.

\subsection{Computations for the Tests}
\label{sec:tests}
\textit{Question 1.}  
Based on the credible region~\eqref{eq:ci_gaussian}, we derive the test statistic
\begin{equation}\label{eq:test_equal}
   q_{\hat{\alpha}} = \left(\f^\star_{\imm} - \map \right)^\intercal 
   \left(\hess_{\im_{1,2}}^{-1} + \hess_{\imm}^{-1} \right)^{-1} 
   \left(  \f^\star_{\imm} - \map \right),
\end{equation}
which follows a chi-squared distribution. 
Therefore, the intersection in \eqref{eq:q1} is null---and, thus, we reject $H_0$---if 
\begin{equation}\label{eq:comparestatistic}
   q_{\hat{\alpha}} > q_\alpha,
\end{equation}
which is a simple comparison between the product of known quantities and the chi-squared quantile of the prescribed minimum confidence level $\alpha$. This is based on considering the random variable resulting from the difference $\frandc-\frand |\imm$. Note that the Hessians in~\eqref{eq:test_equal} are those in \eqref{eq:covariance}. (See Appendix~\ref{sec:implementation} for details about the computational inversion of these matrices.) 
We remark that this test in \eqref{eq:test_equal} is very similar to Wald's test under the conditions listed in Appendix~\ref{sec:lsq}. 

If we restrict the space of interest from the FOV to a subspace $Z \subset \text{FOV}$, then 
$H_0: \map \big|_\patch = \f^\star_{\imm} \big|_\patch$
and the test becomes
\begin{align}\label{eq:test_equal_z}
   \left( \f^\star_{\imm, \patch}- \f^\star_{\im_{1,2}, \patch} \right)^\intercal 
   &\left(\hess_{\imm, \patch}^{-1} + \hess_{\im_{1,2}, \patch}^{-1} \right)^{-1} \\\ 
   &\left( \f^\star_{\imm, \patch}- \f^\star_{\im_{1,2}, \patch} \right) > q_\alpha.
\end{align}
The subspace $\patch$ could be a force patch inside a cell.

\bigskip

\textit{Question 2.} For question 2, our test statistic is given by
\begin{equation}\label{eq:question2statistic}
   q_{\hat{\alpha}} = \left(\b^\star - \map \right)^\intercal 
   \left(\hess_{\im_{1,2}}^{-1} + \mathbf{B} \right)^{-1} 
   \left( \b^\star - \map \right), %
\end{equation}
where $\mathbf{B}$ is the covariance associated with the random variable $\brand$. As in question 1, we reject the null hypothesis when $q_{\hat{\alpha}} > q_\alpha$. Note that if $\mathbf{B}\approx \mathbf{0}$, then the inversion of the Hessian is not needed.

\bigskip

\textit{Question 3.} We reject the null hypothesis of question 3 with a confidence level $\alpha$ when
\begin{equation}\label{eq:question3a}
  \bigwedge_{\f \in I} \left(\big(\f - \map \big)^\intercal 
   \hess_{\im_{1,2}}
   \big( \f - \map \big) > q_\alpha \right),
\end{equation}
or, equivalently, when
\begin{equation}\label{eq:question3b}
  \neg \bigvee_{\f \in I} \left(\big(\f - \map \big)^\intercal 
   \hess_{\im_{1,2}}
   \big( \f - \map \big) < q_\alpha \right),
\end{equation}
where $\bigwedge$ and $\bigvee$ stand for the logical ``and'' and ``or'', respectively. The correponding statistic is thus 
\begin{equation}\label{eq:question3statistic}
q_{\hat{\alpha}}=\min_{\f \in I} \big(\f - \map \big)^\intercal \hess_{\im_{1,2}} \big( \f - \map \big). 
\end{equation}
Details on the set $\inp$ of alternative, inpainted force maps are given in Appendix~\ref{sec:inpainting}.

\bigskip

\textit{Bonus Question (4).} Many other questions are possible. For example, instead of comparing two force patterns, we can compare their magnitudes. Are all adhesion patches generating the same amount of force? This translates into $H_0: ||\map \big|_\patch||_{2}^2 = ||\f^\star_{\imm}\big|_\patch||_{2}^2$ to compare each point spatially. In the Gaussian case, the distributions follow a generalized chi-squared distribution. The distribution is of the same type (with different parameters) if one wants to compare the total force exerted as per $H_0: ||\map \big|_\patch||_{2,\patch}^2 = ||\f^\star_{\imm}\big|_\patch||_{2,\patch}^2$, where these subscripts stand for the norm over $\patch$.

\bigskip

Find a discussion in Appendix~\ref{app:nonlinear} about potential extensions of these tests to problems under other assumptions.

\subsection{Inpainting}\label{sec:inpainting}
To generate an alternative measurement without the force patch, we segment and mask the region of interest. We then fill in this region by solving an interpolation (or inpainting) problem~\cite{aubert_other_2006}. 
More precisely, the new map without the patch~is 
\begin{equation}\label{eq:force_patch}
 \finp(\x; \lambda) = \arg \min_{\f \in F} \, \int_{\x \in \fov \setminus \patch} (\f - \map)^2 + \lambda \reg\{ \f \}, %
\end{equation}
where the optimal parameter $\lambda=\lambda^\star$ can be chosen according to the noise of the measured force map. (Alternatively, one could use other methods such as biharmonic inpainting.) %

Notice that \eqref{eq:force_patch} is a single alternative force map. The test then becomes whether this map belongs to the (smallest) credible region, i.e. we reject the null hypothesis if $\null=\inp=\finp(\x; \lambda^\star) \notin \ci^\alpha_{\frandc}$. 

To make $H_0$ more encompassing, we also studied the possibility of including a whole set of inpainted maps as potential alternatives. 
We propose to generate this set by allowing for not only a single $\lambda$ but a range thereof:
\begin{equation}
   \inp = \{ \finp(\x; \lambda) \, | \, \lambda \in \Lambda  \}.
\end{equation}
In principle, $\Lambda \subset \reals$ should be a discrete set ($|\Lambda|<\infty$) of a few select $\lambda$. Nevertheless, in Appendix~\ref{app:nonlinear}, we discuss the implications of making the set $\inp$ continuous, e.g. through a continuous hyperparameter space such as $\Lambda=[0,1]$.

An alternative to inpainting would be to rerun the reconstruction algorithm while withholding any information about the zone of interest. However, it is unclear how this would impact the reconstruction in other areas, as local forward problems usually become nonlocal when inverted.

\subsection{Statistical Significance vs Practical Significance} \label{sec:practical}
Hypothesis tests are meant to assess \textit{statistical} significance. This does not always translate well into \textit{practical} significance, especially when the null hypothesis is a single-point hypothesis (such as in \eqref{eq:q1} and \eqref{eq:background}), or when the data size is small or large~\cite{bergh_sample_2015,lin_research_2013, chen_roles_2023,colquhoun_reproducibility_2017}. Although this issue is well-known and is common to hypothesis tests across all science, it continues to be overlooked relatively often. (This includes the single-point t-tests that are ubiquitous in biology articles.) In response, some disciplines have opted to establish very stringent significance levels (e.g., genetics or particle physics); whereas others recommend to report an effect-size measure alongside the p-value or, more recently, to turn towards equivalence-test procedures such as the so-called TOST adaptation of t-tests~\cite{lakens_equivalence_2017,halsey_reign_2019}.

If we want to imbue our questions with more practical significance, we propose to consider (ball) sets $\mathrm{B}_\epsilon(\f) \subset \fspace$ of all force maps that are an (Euclidian) distance $\epsilon$ or less from a certain map $\f$ of interest: Instead of testing for $\{\mathbf{0} \}$ in \eqref{eq:q1}, one can test for $\mathrm{B}_\epsilon(\mathbf{0})$; and one can test for $\mathrm{B}_\epsilon(\vb^\star)$ in place of $\{ \vb^\star \}$ in \eqref{eq:background}. In a similar spirit to TOSTs, 
our modified tests allow to inquire about statistical and practical significance in intertwinement. Question 1 is no longer whether the force maps are statistically equal, but whether they are statistically less than $\epsilon$ apart. Notice that this introduces an additional variable $\epsilon$, on top of $\alpha$, that is meant to establish whether a difference is relevant in practice. The choice of $\epsilon$ should be application- or domain-specific and ideally stem from a community-wide consensus.

Under this framework, we compute
\begin{align}
q_{\hat{\alpha}}=\min_{\f \in \mathrm{B}_\epsilon(\mathbf{0})} \left(\f+\f^\star_{\imm} - \map \right)^\intercal 
   &\left(\hess_{\im_{1,2}}^{-1} + \hess_{\imm}^{-1} \right)^{-1} \nonumber \\
   &\left( \f+ \f^\star_{\imm} - \map \right) 
\end{align}
for question 1, where the ball is defined as $\mathrm{B}_\epsilon(\f) := \{ \hat{\f} \in \fspace \, | \, ||\hat{\f} - \f ||_{\text{FOV}, 2}<\epsilon \}$. Alternatively, one could directly check whether the two convex sets (an ellipsoid and a sphere) intersect. This can be achieved via closed-form, convex-optimization, or Monte-Carlo algorithms~\cite{boyle_method_1986,gilitschenski_direct_2014,rabiei_intersection_2021}; the last of which can even quantify the volume of the intersection and, thus, potentially make the test more precise. 
Note that the ball can be substituted with an ellipsoid that allows for a vectorial $\epsilon$.

Together, the covariance, the eigenvector-based visualization of credible regions, and the hypothesis tests can help summarize the uncertainty of the measurements intelligibly. Our intention, however, is not to favor one strategy or the other, but to provide a possible statistical framework to encourage a debate in the community. %

\subsection{Least-Squares Perspective Leads to Wald's Test}\label{sec:lsq}
Here, we show that the interpretation of the intersection that we detailed in Section~\ref{sec:interpretationalpha} coincides with Wald's test in the context of test~\eqref{eq:test_equal} for question 1.

We restrict ourselves to discrete spaces for clarity (see Appendix~\ref{sec:notation}). %
Consider the force $\frand$ as the random variable resulting from
\begin{equation}
 \frand (\interand) \sim \arg \min_{\f \in \fspace} \, \of \{\v; \interand \} + \reg(\f) \text{ subject to } m(\v; \f)=0
\end{equation}
when $\interand$ is a random Gaussian variable modelling image noise with mean $\im_2-\im_1$ and covariance $\mathbf{W}^{-1}$. 
We take the regularization to be $\reg(\f)=\beta ||\mathbf{L}\f||_2^2$. %
We also consider that the PDE model is linear as is the case in many of the problems. Then the PDE model can be written as $\mathbf{M} \v = \f$, where $\mathbf{M}$ is the matrix of the model under the chosen discretization\footnote{In certain PDEs, e.g.~\eqref{eq:elastic}, not only the displacement (or velocity) is present, but also auxiliary variables such as pressure. In such cases, we can incorporate these variables into $\v$ as well to fit the written formalism.}; for example, of finite elements. The minimization problem 
follows the solution
\begin{equation}
    \map = -2\mathbf{C} \mathbf{M}^{-\intercal}\mathbf{O}^\intercal\mathbf{W} (\im_2-\im_1),
\end{equation}
to the equivalent (over-determined) least-squares problem, where
\begin{equation}
   \mathbf{C} = \mathbf{H}^{-1} = \left( 2\mathbf{M}^{-\intercal}\mathbf{O}^\intercal\mathbf{W}\mathbf{O}\mathbf{M}^{-1} + 2\beta\mathbf{L}^\intercal \mathbf{L} \right)^{-1}.
\end{equation}
Since this is a linear transformation of the Gaussian noise modeled by $\interand$, the resulting random variable $\frand$ is a Gaussian with mean $\map$ and covariance $\mathbf{C}$.

Given two different sets of measurements $\im_{2}-\im_{1}$ and $\im_{4}-\im_{3}$ taken under two different conditions, we can consider using Wald's test to question whether the reconstructed forces are similar. We first subtract the two forces and then test if the result is close to zero. Specifically, the test ascertains that
\begin{equation}
(\f^\star_{\imm}-\map-\mathbf{0})^\intercal \mathbf{C}_{\im_{1,2},\im_{3,4}}^{-1}(\f^\star_{\imm}-\map-\mathbf{0})
\end{equation}
follows a chi-squared distribution with $\mathbf{C}_{\im_{1,2},\im_{3,4}}=\mathbf{C}_{\im_{1,2}}+\mathbf{C}_{\im_{3,4}}$. Therefore, the final test is the same as in~\eqref{eq:test_equal}, which we derived from the perspective of the intersections. We remark that an F-test may also be formulated using the perspective of the least-squares problem.

\subsection{Optimization and Algorithm Details}\label{sec:implementation}
Our reconstructions, covariances, credible regions, and tests, all account for the vectorial nature of the fields and for a finite-element basis. We chose to use the finite-element method (FEM)~\cite{alnaes_fenics_2015} because the domains in which the PDEs of the physical model are defined can take arbitrary shapes. For instance, for the TFM example, the force $\f$ is forced to zero when far from the cell (whereas $\v$ is free over the entire $\fov$, see Figure~\ref{fig:framework}). Similarly, for the AN example, the entire system is restricted to the disk, which we take directly as the $\fov$. Details on how to discretize everything with the FEM in similar situations can be found in~\cite{boquet-pujadas_reformulating_2022}. We implemented the FEM with the help of the FEniCS and hiPPYlib libraries~\cite{alnaes_fenics_2015,villa_hippylib_2018}. We generated the meshes using CGAL~\cite{the_cgal_project_cgal_2024}.

We solve the inverse problem~\eqref{eq:inverse} using a multiresolution, gradient-based optimization, in which the model constraint $m(\v; \f)=0$ and its total derivatives are handled with the adjoint method. The result of the inverse problem is the reconstruction or force measurement $\map$. 

We chose the regularization parameter according to Morozov's criterion~\cite{morozov_criteria_1984}, which, for our one-step algorithm, avoids overfitting by recognizing that the error in our data term should be of the order of the image noise. %

\begin{figure}
    \centering
    \includegraphics[width=0.7\linewidth]{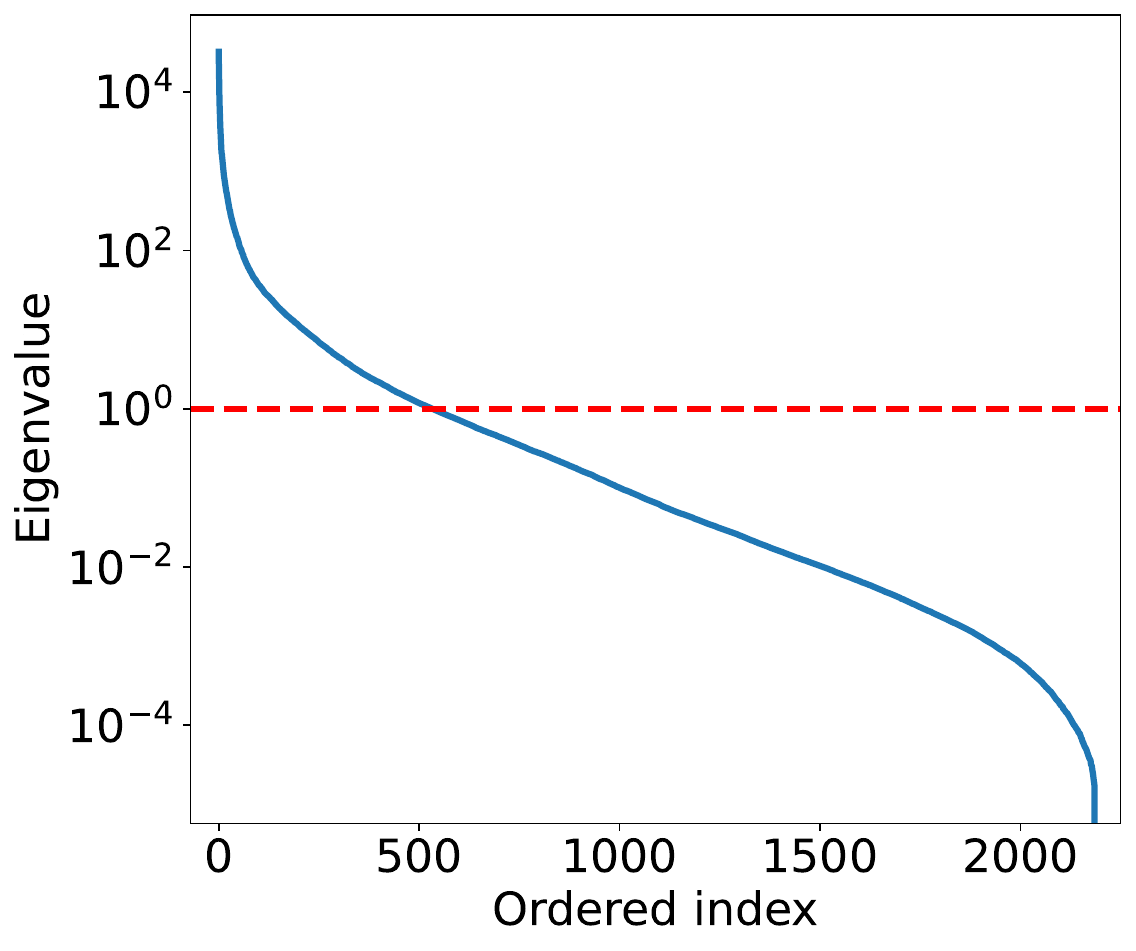}
    \vspace{-5pt}
    \caption{\textbf{Rapid decay of the eigenvalues of the Hessian-based generalized eigenvalue problem in~\eqref{eq:eigenvalue_problem}.} The ordered index is simply the index of the eigenvalues after sorting them.}
    \vspace{-5pt}
    \label{fig:eigenvalues}
\end{figure}

Inverting the Hessian matrices is required for \eqref{eq:covariance} and \eqref{eq:test_equal}, but it is computationally expensive because of their size. Therefore, we invert them by using the low-rank approximation
\begin{equation}\label{eq:low-rank}
\hess^{-1} \approx \mathbf{R}^{-1} - \mathbf{U} \mathbf{\Gamma} \mathbf{U}^\intercal,
\end{equation}
where $\mathbf{R} = 2\beta\mathbf{L}^\intercal \mathbf{L}$. Here, the columns of the $\fdim \times r$ matrix $\mathbf{U}$ are made up of the first $r$ generalized eigenvectors of the problem 
\begin{equation}\label{eq:eigenvalue_problem}
    \left(\hess - \mathbf{R} \right) \f = \gamma_k \mathbf{R} \f
\end{equation}
with respect to $\f$, whereas $\mathbf{\Gamma}$ is an $r \times r$ diagonal matrix made up of the transformation $\gamma_k/(\gamma_k+1)$ of the $r$ first corresponding eigenvalues $\gamma_k$, where $r$ is the desired rank~\cite{borzi_computational_2011,spantini_optimal_2015,boquet-pujadas_reformulating_2022}. The error of approximation~\eqref{eq:low-rank} is dominated by %
$\gamma_r/(\gamma_r+1)$.
The spectrum of the image-data part $\left(\hess - \mathbf{R} \right)$ of the Hessian decays rapidly, as only a small subspace of the parameters is informed by the images acquired (Figure~\ref{fig:eigenvalues}). This allows us to choose a relatively small rank $r \ll h$ such that $\gamma_r \ll 1$, ensuring a good approximation of the matrix at a reduced dimension, which accelerates computations. We compute this low-rank approximation by passing random vectors through the action of the matrix (i.e., without explicitly constructing it) as per a randomized algorithm for generalized Hermitian eigenvalue problems. In particular, we use a generalized double-pass algorithm~\cite{saibaba_randomized_2016,halko_finding_2011}.

With the map $\map$ and the approximation of the inverse Hessian $\hess^{-1}$, one can perform the hypothesis tests as outlined in Appendix~\ref{sec:tests}. The entire pipeline is summarized in Algorithm~\ref{algo:pipeline}.

\begin{algorithm}
\caption{Statistical Assessment of Cell Forces}\label{algo:pipeline}
	\KwData{A pair of acquired images $\im_{1,2}$ (or two pairs, e.g., $\im_{3,4}$), a null hypothesis $H_0$ with its associated set $N$, and a prescribed minimum confidence level $\alpha$. 
 }
	\KwResult{Whether the null hypothesis is rejected in favor of $H_1$.}
 
	Find the most probable force map $\map$ given the acquired images by solving~\eqref{eq:inverse} via optimization\; 
 
	Find the covariance of the measurement error by inverting ($\hess^{-1}$) the Hessian~\eqref{eq:hessian} using the low-rank approximation~\eqref{eq:low-rank}\;

   Visualize the statistical variability of the force maps through the most relevant directions of the credible regions using~\eqref{eq:eigen_regions}\; 
 
	Compute the
 statistic of the corresponding hypothesis (or equivalence) test through~\eqref{eq:test_equal}, \eqref{eq:question2statistic}, \eqref{eq:question3statistic} and use the cumulative distribution function (CDF) of the chi-squared function to obtain the p-value $\hat{\alpha}$\;

 Decide whether to reject the hypothesis by comparing $\hat{\alpha}$ to the prescribed confidence level $\alpha$ (or by comparing the quantiles $q_{\hat{\alpha}}$, $q_\alpha$ directly as per~\eqref{eq:comparestatistic})\;
 \end{algorithm}

 To compute the eigenvalues of $\hess$ (or of the covariance $\hess^{-1}$, since it is a symmetric (semi) positive-definite matrix) that we use to visualize the credible regions via~\eqref{eq:eigen_regions}, we employ a randomized double-pass algorithm for (non-generalized) Hermitian eigenvalue problems.

\subsection{Experiment Details}\label{sec:experiment_details}

To illustrate the framework in different situations while remaining concise, we selected two types of experiments: one where the system is a solid, and the other where it is a fluid.
\begin{itemize}
    \item The solid system follows the compressible linearly elastic model $m(\v;\f)=0$~\eqref{eq:elastic}
    in the context of TFM, where cells exert tractions on the substrate. Here, $\v$ stands for the displacement. Two images are taken for each TFM measurement: one when the substrate is at rest, and one while it is being actively deformed by the cell.
    \item The fluid system follows the Stokes model in~\cite[Equation 3]{boquet-pujadas_inverse_2023} in the context of AN hydrodynamics such as in microtubule-kinesin mixtures in suspension. Here, $\v$ acts as a~velocity. In this case, images are taken as the system evolves.
\end{itemize}
In both cases, we use Dirichlet boundary conditions, which can be guessed as part of the algorithm by adding them to the ``force'' term in~\eqref{eq:inverse}. Note that the force measurements for both systems are given in relative units of $\f/\mu$, where $\mu$ stands for the shear modulus in the TFM example, and for the viscosity in the AN example. (We made this choice because the viscosity in AN systems is difficult to measure.) The force density $\f$ itself has units of\unit{}{[N .  \meter^{-2}]} in 2D, whereas the 2D shear modulus $\mu$ has units of\unit{}{[Pa . \meter]}, and the 2D viscosity $\mu$ of\unit{}{[Pa . \meter . \second]}. (To switch to 3D one can use $\mu_{\text{3D}}=\mu/h$ and $\f_{\text{3D}}=\f/h$, where $h$ is the relevant thickness of the material.) As a result, the units of $\f/\mu$ are \unit{}{[\meter^{-1}]} and \unit{}{[(\meter . \second)^{-1}]} for the respective relative forces. On the other hand, the displacement (TFM) and the velocity (AN) $\v$ have units of\unit{}{[ \meter]} and\unit{}{[\meter /\second]} because the pixel size and the frame rate are readily available for both examples. 

Here is a description of the data for the two types of measurement experiments:
\begin{itemize}
    \item For TFM, the experiments were run on data taken in the context of the work in~\cite{mittal_high-resolution_2021,mittal_myosin-independent_2024}, as well as of the work in~\cite{sune-aunon_full_2017,jorge-penas_free_2015}. Both datasets were kindly shared by the respective authors upon request (see Acknowledgements). The TFM example in the figures shows a Chinese-hamster ovary cell from the CHO-K1 line (expressing Lifeact-GFP) that is adhering to a polyacrylamide substrate ($E=\unit{5}{\kilo \pascal}$ and $\nu=0.45$ for the Young's modulus and Poisson's ratio, respectively) seeded with fluorescent polystyrene beads of \unit{0.2}{\micro \meter} in diameter (with Alexa Fluor 647 for far-red emission)~\cite{aguilar-cuenca_nuevos_2017}. The images were acquired with a confocal microscope (Leica TCS SP5 II). The image of the substrate at rest was taken after lysing the cell. 
    \item For the AN systems, the data is taken directly from the dataset in~\cite{opathalage_self-organized_2019}. The AN system in the figures is a mixture of microtubules (labeled with Alexa Fluor 647), kinesin clusters (bound by tetrameric streptavidin), and a depleting agent~\cite{opathalage_self-organized_2019}. The images were acquired with a wide-field fluorescence microscope (Nikon Eclipse Ti-E).
\end{itemize}

The parameters and details used for the experiments are listed here:
\begin{itemize}
    \item %
    To compute the standard deviation of the TFM image data, we exploited regions of the images that contained virtually no signal. We found a \unit{12}{\%} error for the brightness with respect to its mean, which is then compounded by the fact that our computations are based on the difference between two images. For question 2, the background was assumed to be distributed as a Gaussian characterized by these parameters. The optimal regularization parameter $\beta^\star$ was chosen among several $\beta$ based on Morozov's criterion to match the level of image noise. The inpainting set for question 3 was then defined as $\Lambda=\{10^{k} \beta^\star \, | \, k \in \{-2,-1,\dots,2\} \}$.
Unless specified otherwise, we chose a significance level of $\alpha=\pv$ for the tests. To define $\mathrm{K}$, we used an active-contours method~\cite{kass_snakes_1988,manich_protocol_2020} to segment the cell and dilated the resulting polygon away from it. We took the whole image as the $\fov$ and meshed it in consideration of the dilated boundary $\partial \mathrm{K}$. We used first-degree triangular finite elements to discretize the TFM example, resulting in $\fdim = 20,402$ degrees of freedom for the force.

\item The experiments for the AN systems were conducted similarly.
We measured the relative standard deviation of the image noise at \unit{15}{\%}, which is then compounded. We used an active-contours method to segment the disk, which we set as the $\fov$, instead of using the whole image. We then meshed the disk with finer elements near the boundary and used a basis of first-degree triangular elements for everything but the velocity $\v$ (second degree). This resulted in $\fdim = 8,402$ degrees of freedom. For question 1, we applied a rotation to one of the measurements and projected it onto the finite-element basis of the other. (We also had to project the Hessian.) This was done to align the defect being propelled forward, which happens at different points of the circular boundary for different periods of the dynamics.
\end{itemize}

For the background noise $\brand$ in question 2, we use the same relative standard deviation percentage as for the image noise, but calculated with respect to the reconstructed measurement. We set the mean of the background $\vb^\star(\x)$ to be a zero vector field for both components. When applicable, the $\epsilon$ for the practical ball is taken as \unit{5}{\%} of the median of the measurement.

\subsection{Prospective Extensions}\label{app:nonlinear}
\subsubsection{Extension to Non-Gaussian, Non-Convex, Non-Linear, or Non-Differentiable Terms}

Certain properties of the posterior can be very helpful for its characterization. Two of them are log-concavity and Gaussianity. The probability density function of a posterior $\den_{\frandc} \propto \exp(-\mathcal{D}(\f)-\reg(\f))$ is log-concave if, and only if, the variational energy $\mathcal{D}(\f)+\reg(\f)$ within it is convex. If, in addition, both terms are Gaussian, then the posterior is Gaussian. Another important property is whether the energy terms are differentiable or, at least, have a conducive proximal operator. 

A Gaussian posterior normally originates from a Gaussian likelihood (i.e. $\mathcal{D}$ modelling Gaussian noise combined with a linear operator such as $\mathbf{O}\mathbf{M}^{-1}$) and a Gaussian prior (e.g., $\mathcal{R}$ being the combination of $L^2$ and a linear operator such as $\mathbf{L}$). This is the most common situation in mechanobiology problems. Sometimes either of these terms is convex but not differentiable; for example, total-variation (TV) regularization is not differentiable (unless an epsilon term is added to the argument), but has a well-studied proximal operator~\cite{beck_fast_2009}. This is also true for its Hessian counterpart (HTV)~\cite{lefkimmiatis_hessian_2013,boquet-pujadas_sensitivity-aware_2024,pourya_box-spline_2024}. In typical inverse problems, both terms are convex. Possible sources of non-convexity in \eqref{eq:covariance}-\eqref{eq:hessian}, \eqref{eq:distributedas_gaussian}-\eqref{eq:ci_gaussian} could be non-linear physical models in $m=0$ (unlike the mechanobiology methods in the introduction), Poisson-dominant noise, or big deformations in the optical-flow term (so that the linearization is not accurate). 
In the latter case, however, \eqref{eq:covariance}-\eqref{eq:hessian} (and thus the rest of equations) can still constitute a good approximation~\cite{boquet-pujadas_reformulating_2022} because they capture the second-order information (Laplacian approximation).

\paragraph{The MAP} Regardless of convexity or Gaussianity, the maximum a posteriori (MAP) point $\map$ of the posterior $\den_{\frandc}$ is always the result of solving the problem in~\eqref{eq:inverse}. For convex problems, one can use gradient or proximal descent algorithms to such end; whereas, solving non-convex problems requires more complex approaches such as multiresolution or convexification techniques.

\paragraph{The Credible Regions}

In the convex setting, the credible regions resulting from the posterior are convex sets. Among them, the highest posterior density (HPD) region is the one with the smallest volume for a given confidence level. These regions are delimited by the level sets of the convex function and thus can be characterized by a single value $c_\alpha$ as $\{ \f \in \fspace \, | \, \mathcal{D}(\f)+\reg(\f) < c_\alpha \}$. %
The characterization of these regions is useful for testing. 

It is easier to work out $c_\alpha$ if the posterior is Gaussian. In such a case, the posterior is completely characterized by the mean (which is also the MAP) and the covariance (see $\eqref{eq:covariance}$), and the credible regions are ellipsoids (see \eqref{eq:ci_gaussian}).
The characterization of $c_\alpha$ is not as straightforward for arbitrary $\mathcal{D}$ or $\mathcal{R}$~\cite{javanmard_confidence_2013,bobkov_concentration_2011}. 
In such cases, Markov Chain Monte Carlo (MCMC) algorithms can be used to sample from the posterior $\den_{\frandc}(\f)$ in order to build the credible regions. MCMC algorithms can be sped up greatly by exploiting information from the gradient or the Hessian of $\mathcal{D}$ and $\mathcal{R}$. If either term is not differentiable, %
one can exploit the proximal operator instead.%

MCMC algorithms are slow when the space $\fspace$ is high dimensional. Here we comment on two alternatives. The first one is to take a Laplacian approximation, i.e. to approximate the posterior as a Gaussian. In such case, \eqref{eq:covariance}-\eqref{eq:hessian} apply. This provides fast and rather efficient credible regions of the form of \eqref{eq:ci_gaussian} based on the covariance and mean of the actual density, but without real guarantees. The second alternative hinges on our preference for type I errors: there exist approximations that overestimate the value of $c_\alpha$ but only require knowing the maximum a posteriori (MAP) of the density~\cite{bobkov_concentration_2011}. Although we expect these to get worse as the dimension increases, they could be used to vastly increase the testing speed at the price of requiring more evidence to reject the null hypothesis (type I error). It is worth noting that sampling convex energies effectively is significantly easier (and faster) because the underlying distributions are unimodal. The gradient-based sampling that accelerates the exploration of the sampling space of log-concave distributions is less efficient when dealing with the multi-modality of non-convex energies.

\subsubsection{Large Hypothesis Sets} Another potential extension is to consider null-hypothesis sets that have a non-finite (or very large) number of elements. In this case, it becomes unfeasible to test each of the candidates in the set. This is less of a problem if one expects that the sets do intersect because, once we find a single $\f$ such that $\f\in \null$ and $\f \in \ci^\alpha_{\frand|\im_{1,2}}$, we can stop. Another perspective is that, if we work with an $\null$ set that is convex (or construct it to be so), then the distance between sets can be computed with Dykstra's algorithm~\cite{boyle_method_1986,bauschke_projection_1996} of alternating projections because $\ci^\alpha_{\frand|\im_{1,2}}$ is convex, too. By choosing a tolerance for the distance, this approach could provide a principled method of assessing whether the sets intersect for the test.

Notice that other testing alternatives than those presented here, like likelihood-ratio tests or Bayes' factors (the ratio of probabilities of any two hypotheses weighted by those of the priors thereunder)~\cite{good_bayesnon-bayes_1992}, can be readily computed within this Bayesian framework, too. These alternatives can also be used to set hyperparameters.

\subsubsection{Marginalization} Note that one could also use Bayes rule to marginalize the error incurred by using the model $m(\v;\f)=0$ instead of a potentially more complex one, or that incurred by the fact that the parameters in $m(\v;\f)=0$ (e.g., the Young's modulus) are measured experimentally. The simplest approach to this is to marginalize them as additive Gaussians and incorporate them into $c$ and $\mathbf{W}^{-1}$. In such a case, the mean and covariance can be characterized by running simulated experiments.

\subsection{Notation}\label{sec:notation}
While the main text is written using functions such as $\f(\x)$ for clarity, the details exposed in the Appendix are considerably simpler using (discretized) vectors instead. (E.g., it allows working with multivariate Gaussians instead of Gaussian processes and probability density function[al]s.) For this, consider that we choose $\fspace = \reals^{\fdim}$ in the main text, and that the notation $\f(\x)$ stands for the indexing of the vector $\f$ with an index $\x \in \naturals^d$ (think $\f[\x]$), where $d \in \{2,3\}$ is the dimension of the 2D or 3D space of the problem. Then, $\fdim=d\times n_1 \times \dots \times n_d$ is the size of the vector fields $\f$ or $\v$ over the image (2D) or volume (3D) of size $n_1 \times \dots \times n_d$. The elements of $\f$ can be the values of the force map at each point of the image/volume, or the coefficients of a more elaborated discretization scheme such as those in the finite-element method (FEM).  Domain sets such as $\text{FOV}$ or $\patch$ become index subsets of $\naturals^d$ instead of subsets of $\reals^d$; and any integral turns into a sum over said indices, which also means that the $L^2(\patch)$ norm is replaced by the $\ell^2(\patch)$ norm. From this perspective, $\im_{\bullet}$ becomes a vector in $\reals^{n_1 \times \dots \times n_d}$, too, even if it is not bold like the vector fields~are.

We use a tilde to distinguish random variables (e.g., $\frand$) from realization thereof (e.g., $\f$).

\clearpage
\bibliographystyle{IEEEtran}
\bibliography{IEEEabrv,references,sample}

\begin{thebibliography}{10}
\providecommand{\url}[1]{#1}
\csname url@samestyle\endcsname
\providecommand{\newblock}{\relax}
\providecommand{\bibinfo}[2]{#2}
\providecommand{\BIBentrySTDinterwordspacing}{\spaceskip=0pt\relax}
\providecommand{\BIBentryALTinterwordstretchfactor}{4}
\providecommand{\BIBentryALTinterwordspacing}{\spaceskip=\fontdimen2\font plus
\BIBentryALTinterwordstretchfactor\fontdimen3\font minus \fontdimen4\font\relax}
\providecommand{\BIBforeignlanguage}[2]{{%
\expandafter\ifx\csname l@#1\endcsname\relax
\typeout{** WARNING: IEEEtran.bst: No hyphenation pattern has been}%
\typeout{** loaded for the language `#1'. Using the pattern for}%
\typeout{** the default language instead.}%
\else
\language=\csname l@#1\endcsname
\fi
#2}}
\providecommand{\BIBdecl}{\relax}
\BIBdecl

\bibitem{iskratsch_appreciating_2014}
T.~Iskratsch, H.~Wolfenson, and M.~P. Sheetz, ``\BIBforeignlanguage{en}{Appreciating force and shape — the rise of mechanotransduction in cell biology},'' \emph{\BIBforeignlanguage{en}{Nature Reviews Molecular Cell Biology}}, vol.~15, no.~12, pp. 825--833, Dec. 2014.

\bibitem{editorial_forces_2017}
Editorial, ``\BIBforeignlanguage{en}{Forces in cell biology},'' \emph{\BIBforeignlanguage{en}{Nature Cell Biology}}, vol.~19, no.~6, pp. 579--579, Jun. 2017.

\bibitem{mongera_fluid--solid_2018}
A.~Mongera \emph{et~al.}, ``\BIBforeignlanguage{en}{A fluid-to-solid jamming transition underlies vertebrate body axis elongation},'' \emph{\BIBforeignlanguage{en}{Nature}}, vol. 561, no. 7723, pp. 401--405, Sep. 2018.

\bibitem{romani_extracellular_2019}
P.~Romani \emph{et~al.}, ``\BIBforeignlanguage{en}{Extracellular matrix mechanical cues regulate lipid metabolism through {Lipin}-1 and {SREBP}},'' \emph{\BIBforeignlanguage{en}{Nature Cell Biology}}, vol.~21, no.~3, pp. 338--347, Mar. 2019.

\bibitem{tajik_transcription_2016}
A.~Tajik \emph{et~al.}, ``\BIBforeignlanguage{en}{Transcription upregulation via force-induced direct stretching of chromatin},'' \emph{\BIBforeignlanguage{en}{Nature Materials}}, vol.~15, no.~12, pp. 1287--1296, Dec. 2016.

\bibitem{andreu_mechanical_2022}
I.~Andreu \emph{et~al.}, ``\BIBforeignlanguage{en}{Mechanical force application to the nucleus regulates nucleocytoplasmic transport},'' \emph{\BIBforeignlanguage{en}{Nature Cell Biology}}, vol.~24, no.~6, pp. 896--905, Jun. 2022.

\bibitem{marx_may_2019}
V.~Marx, ``\BIBforeignlanguage{en}{May mechanobiology work forcefully for you},'' \emph{\BIBforeignlanguage{en}{Nature Methods}}, vol.~16, no.~11, pp. 1083--1086, Nov. 2019.

\bibitem{boquet-pujadas_bioimage_2021}
A.~Boquet-Pujadas, J.-C. Olivo-Marin, and N.~Guillén, ``\BIBforeignlanguage{English}{Bioimage {Analysis} and {Cell} {Motility}},'' \emph{\BIBforeignlanguage{English}{Patterns}}, vol.~2, no.~1, Jan. 2021.

\bibitem{roca-cusachs_quantifying_2017}
P.~Roca-Cusachs, V.~Conte, and X.~Trepat, ``\BIBforeignlanguage{en}{Quantifying forces in cell biology},'' \emph{\BIBforeignlanguage{en}{Nature Cell Biology}}, vol.~19, no.~7, pp. 742--751, Jul. 2017.

\bibitem{polacheck_measuring_2016}
W.~J. Polacheck and C.~S. Chen, ``\BIBforeignlanguage{en}{Measuring cell-generated forces: a guide to the available tools},'' \emph{\BIBforeignlanguage{en}{Nature Methods}}, vol.~13, no.~5, pp. 415--423, May 2016.

\bibitem{legant_measurement_2010}
W.~R. Legant \emph{et~al.}, ``\BIBforeignlanguage{en}{Measurement of mechanical tractions exerted by cells in three-dimensional matrices},'' \emph{\BIBforeignlanguage{en}{Nature Methods}}, vol.~7, no.~12, pp. 969--971, Dec. 2010.

\bibitem{han_traction_2015}
S.~J. Han, Y.~Oak, A.~Groisman, and G.~Danuser, ``\BIBforeignlanguage{en}{Traction microscopy to identify force modulation in subresolution adhesions},'' \emph{\BIBforeignlanguage{en}{Nature Methods}}, vol.~12, no.~7, pp. 653--656, Jul. 2015.

\bibitem{steinwachs_three-dimensional_2016}
J.~Steinwachs \emph{et~al.}, ``\BIBforeignlanguage{en}{Three-dimensional force microscopy of cells in biopolymer networks},'' \emph{\BIBforeignlanguage{en}{Nature Methods}}, vol.~13, no.~2, pp. 171--176, Feb. 2016.

\bibitem{campas_quantifying_2014}
O.~Campàs \emph{et~al.}, ``\BIBforeignlanguage{en}{Quantifying cell-generated mechanical forces within living embryonic tissues},'' \emph{\BIBforeignlanguage{en}{Nature Methods}}, vol.~11, no.~2, pp. 183--189, Feb. 2014.

\bibitem{ellis_curvature-induced_2018}
P.~W. Ellis \emph{et~al.}, ``\BIBforeignlanguage{en}{Curvature-induced defect unbinding and dynamics in active nematic toroids},'' \emph{\BIBforeignlanguage{en}{Nature Physics}}, vol.~14, no.~1, pp. 85--90, Jan. 2018.

\bibitem{boquet-pujadas_inverse_2023}
A.~Boquet-Pujadas, J.~Hardouïn, J.~Wen, J.~Ignés-Mullol, and F.~Sagués, ``Inverse {Measurements} in {Active} {Nematics},'' Dec. 2023, arXiv: 2312.15553 [cond-mat, physics:physics].

\bibitem{blakely_dna-based_2014}
B.~L. Blakely \emph{et~al.}, ``\BIBforeignlanguage{en}{A {DNA}-based molecular probe for optically reporting cellular traction forces},'' \emph{\BIBforeignlanguage{en}{Nature Methods}}, vol.~11, no.~12, pp. 1229--1232, Dec. 2014.

\bibitem{serwane_vivo_2017}
F.~Serwane \emph{et~al.}, ``\BIBforeignlanguage{en}{In vivo quantification of spatially varying mechanical properties in developing tissues},'' \emph{\BIBforeignlanguage{en}{Nature Methods}}, vol.~14, no.~2, pp. 181--186, Feb. 2017.

\bibitem{scarcelli_noncontact_2015}
G.~Scarcelli \emph{et~al.}, ``\BIBforeignlanguage{en}{Noncontact three-dimensional mapping of intracellular hydromechanical properties by {Brillouin} microscopy},'' \emph{\BIBforeignlanguage{en}{Nature Methods}}, vol.~12, no.~12, pp. 1132--1134, Dec. 2015.

\bibitem{prevedel_brillouin_2019}
R.~Prevedel, A.~Diz-Muñoz, G.~Ruocco, and G.~Antonacci, ``\BIBforeignlanguage{en}{Brillouin microscopy: an emerging tool for mechanobiology},'' \emph{\BIBforeignlanguage{en}{Nature Methods}}, vol.~16, no.~10, pp. 969--977, Oct. 2019.

\bibitem{ghosh_image-based_2021}
S.~Ghosh, V.~C. Cuevas, B.~Seelbinder, and C.~P. Neu, ``\BIBforeignlanguage{en}{Image-{Based} {Elastography} of {Heterochromatin} and {Euchromatin} {Domains} in the {Deforming} {Cell} {Nucleus}},'' \emph{\BIBforeignlanguage{en}{Small}}, vol.~17, no.~5, p. 2006109, 2021.

\bibitem{kesenci_estimation_2024}
Y.~Kesenci, A.~Boquet-Pujadas, M.~Unser, and J.-C. Olivo-Marin, ``Estimation of {Stiffness} {Maps} in {Deforming} {Cells} {Through} {Optical} {Flow} with {Bounded} {Curvature},'' \emph{IEEE Transactions on Medical Imaging}, pp. 1--1, 2024.

\bibitem{gomez-martinez_silicon_2013}
R.~Gómez-Martínez \emph{et~al.}, ``\BIBforeignlanguage{en}{Silicon chips detect intracellular pressure changes in living cells},'' \emph{\BIBforeignlanguage{en}{Nature Nanotechnology}}, vol.~8, no.~7, pp. 517--521, Jul. 2013.

\bibitem{boquet-pujadas_bioflow_2017}
A.~Boquet-Pujadas \emph{et~al.}, ``\BIBforeignlanguage{en}{{BioFlow}: a {Non}-{Invasive}, {Image}-{Based} {Method} to {Measure} {Speed}, {Pressure} and {Forces} inside {Living} {Cells}},'' \emph{\BIBforeignlanguage{en}{Scientific Reports}}, vol.~7, no.~1, p. 9178, Aug. 2017.

\bibitem{mittasch_non-invasive_2018}
M.~Mittasch \emph{et~al.}, ``\BIBforeignlanguage{en}{Non-invasive perturbations of intracellular flow reveal physical principles of cell organization},'' \emph{\BIBforeignlanguage{en}{Nature Cell Biology}}, vol.~20, no.~3, pp. 344--351, Mar. 2018.

\bibitem{klughammer_cytoplasmic_2018}
N.~Klughammer \emph{et~al.}, ``\BIBforeignlanguage{en}{Cytoplasmic flows in starfish oocytes are fully determined by cortical contractions},'' \emph{\BIBforeignlanguage{en}{PLOS Computational Biology}}, vol.~14, no.~11, p. e1006588, Nov. 2018.

\bibitem{tambe_collective_2011}
D.~T. Tambe \emph{et~al.}, ``\BIBforeignlanguage{en}{Collective cell guidance by cooperative intercellular forces},'' \emph{\BIBforeignlanguage{en}{Nature Materials}}, vol.~10, no.~6, pp. 469--475, Jun. 2011.

\bibitem{boquet-pujadas_4d_2022}
A.~Boquet-Pujadas \emph{et~al.}, ``{4D} {Live} {Imaging} and {Computational} {Modeling} of a {Functional} {Gut}-on-a-{Chip} {Evaluate} how {Peristalsis} {Facilitates} {Enteric} {Pathogen} {Invasion},'' \emph{Science Advances}, vol.~8, no.~42, p. eabo5767, Oct. 2022.

\bibitem{kim_embryonic_2021}
S.~Kim, M.~Pochitaloff, G.~A. Stooke-Vaughan, and O.~Campàs, ``\BIBforeignlanguage{en}{Embryonic tissues as active foams},'' \emph{\BIBforeignlanguage{en}{Nature Physics}}, vol.~17, no.~7, pp. 859--866, Jul. 2021.

\bibitem{seelbinder_nuclear_2021}
B.~Seelbinder \emph{et~al.}, ``\BIBforeignlanguage{en}{Nuclear deformation guides chromatin reorganization in cardiac development and disease},'' \emph{\BIBforeignlanguage{en}{Nature Biomedical Engineering}}, vol.~5, no.~12, pp. 1500--1516, Dec. 2021.

\bibitem{lv_active_2024}
J.-Q. Lv \emph{et~al.}, ``\BIBforeignlanguage{en}{Active hole formation in epithelioid tissues},'' \emph{\BIBforeignlanguage{en}{Nature Physics}}, vol.~20, no.~8, pp. 1313--1323, Aug. 2024.

\bibitem{head_spontaneous_2024}
L.~C. Head \emph{et~al.}, ``\BIBforeignlanguage{en}{Spontaneous self-constraint in active nematic flows},'' \emph{\BIBforeignlanguage{en}{Nature Physics}}, vol.~20, no.~3, pp. 492--500, Mar. 2024.

\bibitem{pallares_stiffness-dependent_2023}
M.~E. Pallarès \emph{et~al.}, ``\BIBforeignlanguage{en}{Stiffness-dependent active wetting enables optimal collective cell durotaxis},'' \emph{\BIBforeignlanguage{en}{Nature Physics}}, vol.~19, no.~2, pp. 279--289, Feb. 2023.

\bibitem{holenstein_high-resolution_2017}
C.~N. Holenstein, U.~Silvan, and J.~G. Snedeker, ``\BIBforeignlanguage{en}{High-resolution traction force microscopy on small focal adhesions - improved accuracy through optimal marker distribution and optical flow tracking},'' \emph{\BIBforeignlanguage{en}{Scientific Reports}}, vol.~7, no.~1, p. 41633, Feb. 2017.

\bibitem{vaux_replicates_2012}
D.~L. Vaux, F.~Fidler, and G.~Cumming, ``Replicates and repeats—what is the difference and is it significant?'' \emph{EMBO reports}, vol.~13, no.~4, pp. 291--296, Apr. 2012.

\bibitem{waters_accuracy_2009}
J.~C. Waters, ``Accuracy and precision in quantitative fluorescence microscopy,'' \emph{Journal of Cell Biology}, vol. 185, no.~7, pp. 1135--1148, Jun. 2009.

\bibitem{antun_instabilities_2020}
V.~Antun, F.~Renna, C.~Poon, B.~Adcock, and A.~C. Hansen, ``On instabilities of deep learning in image reconstruction and the potential costs of {AI},'' \emph{Proceedings of the National Academy of Sciences}, vol. 117, no.~48, pp. 30\,088--30\,095, Dec. 2020.

\bibitem{boquet-pujadas_reformulating_2022}
A.~Boquet-Pujadas and J.-C. Olivo-Marin, ``Reformulating {Optical} {Flow} to {Solve} {Image}-{Based} {Inverse} {Problems} and {Quantify} {Uncertainty},'' \emph{IEEE Transactions on Pattern Analysis and Machine Intelligence}, pp. 1--16, 2022.

\bibitem{boquet-pujadas_multiple_2018}
------, ``\BIBforeignlanguage{en}{Multiple {Variational} {Image} {Assimilation} for {Accessible} {Micro}-{Elastography}},'' \emph{\BIBforeignlanguage{en}{Journal of Physics: Conference Series}}, vol. 1131, no.~1, p. 012014, Nov. 2018.

\bibitem{robert_bayesian_2007}
C.~P. Robert, \emph{\BIBforeignlanguage{en}{The {Bayesian} {Choice}}}, ser. Springer {Texts} in {Statistics}.\hskip 1em plus 0.5em minus 0.4em\relax New York, NY: Springer, 2007.

\bibitem{casella_statistical_2002}
G.~Casella and R.~Berger, \emph{Statistical {Inference}}, 2nd~ed.\hskip 1em plus 0.5em minus 0.4em\relax New York: Chapman and Hall/CRC, 2002.

\bibitem{shi_reconnecting_2021}
H.~Shi and G.~Yin, ``Reconnecting p-{Value} and {Posterior} {Probability} {Under} {One}- and {Two}-{Sided} {Tests},'' \emph{The American Statistician}, vol.~75, no.~3, pp. 265--275, Jul. 2021.

\bibitem{chen_roles_2023}
O.~Y. Chén \emph{et~al.}, ``The roles, challenges, and merits of the p value,'' \emph{Patterns}, vol.~4, no.~12, p. 100878, Dec. 2023.

\bibitem{sune-aunon_full_2017}
A.~Suñé-Auñón \emph{et~al.}, ``\BIBforeignlanguage{en}{Full {L1}-regularized {Traction} {Force} {Microscopy} over whole cells},'' \emph{\BIBforeignlanguage{en}{BMC Bioinformatics}}, vol.~18, no.~1, p. 365, Aug. 2017.

\bibitem{jorge-penas_free_2015}
A.~Jorge-Peñas \emph{et~al.}, ``\BIBforeignlanguage{en}{Free {Form} {Deformation}–{Based} {Image} {Registration} {Improves} {Accuracy} of {Traction} {Force} {Microscopy}},'' \emph{\BIBforeignlanguage{en}{PLOS ONE}}, vol.~10, no.~12, p. e0144184, Dec. 2015.

\bibitem{opathalage_self-organized_2019}
A.~Opathalage \emph{et~al.}, ``Self-organized dynamics and the transition to turbulence of confined active nematics,'' \emph{Proceedings of the National Academy of Sciences}, vol. 116, no.~11, pp. 4788--4797, Mar. 2019.

\bibitem{schermelleh_super-resolution_2019}
L.~Schermelleh \emph{et~al.}, ``\BIBforeignlanguage{en}{Super-resolution microscopy demystified},'' \emph{\BIBforeignlanguage{en}{Nature Cell Biology}}, vol.~21, no.~1, pp. 72--84, Jan. 2019.

\bibitem{hockmann_analysis_2024}
M.~Hockmann, ``Analysis of the sparse super resolution limit using the {Cramér}-{Rao} lower bound,'' \emph{IEEE Transactions on Information Theory}, pp. 1--1, 2024.

\bibitem{doyon_first_2023}
V.~Doyon \emph{et~al.}, ``\BIBforeignlanguage{en}{First {PET} {Investigation} of the {Human} {Brain} at 2 µ{L} {Resolution} with the {Ultra}-{High}-{Resolution} ({UHR}) scanner},'' \emph{\BIBforeignlanguage{en}{Journal of Nuclear Medicine}}, vol.~64, no. supplement 1, pp. P726--P726, Jun. 2023.

\bibitem{boquet-pujadas_silicon-pixel_2024}
A.~Boquet-Pujadas \emph{et~al.}, ``A {Silicon}-{Pixel} {Paradigm} for {PET},'' \emph{IEEE Transactions on Radiation and Plasma Medical Sciences}, pp. 1--1, 2024.

\bibitem{zhang_fast_2023}
Y.~Zhang \emph{et~al.}, ``\BIBforeignlanguage{en}{Fast and sensitive {GCaMP} calcium indicators for imaging neural populations},'' \emph{\BIBforeignlanguage{en}{Nature}}, vol. 615, no. 7954, pp. 884--891, Mar. 2023.

\bibitem{pham_deep-prior_2024}
T.-a. Pham, A.~Boquet-Pujadas, S.~Mondal, M.~Unser, and G.~Barbastathis, ``\BIBforeignlanguage{en}{Deep-prior {ODEs} augment fluorescence imaging with chemical sensors},'' \emph{\BIBforeignlanguage{en}{Nature Communications}}, vol.~15, no.~1, p. 9172, Oct. 2024.

\bibitem{the_event_horizon_telescope_collaboration_first_2019}
{The Event Horizon Telescope Collaboration}, ``First {M87} {Event} {Horizon} {Telescope} {Results}. {IV}. {Imaging} the {Central} {Supermassive} {Black} {Hole},'' \emph{The Astrophysical Journal Letters}, vol. 875, no.~1, p.~L4, Apr. 2019.

\bibitem{papenberg_highly_2006}
N.~Papenberg, A.~Bruhn, T.~Brox, S.~Didas, and J.~Weickert, ``\BIBforeignlanguage{en}{Highly {Accurate} {Optic} {Flow} {Computation} with {Theoretically} {Justified} {Warping}},'' \emph{\BIBforeignlanguage{en}{International Journal of Computer Vision}}, vol.~67, no.~2, pp. 141--158, Apr. 2006.

\bibitem{berger_testing_1987}
J.~O. Berger and T.~Sellke, ``Testing a {Point} {Null} {Hypothesis}: {The} {Irreconcilability} of {P} {Values} and {Evidence},'' \emph{Journal of the American Statistical Association}, vol.~82, no. 397, pp. 112--122, 1987.

\bibitem{casella_reconciling_1987}
G.~Casella and R.~L. Berger, ``Reconciling {Bayesian} and {Frequentist} {Evidence} in the {One}-{Sided} {Testing} {Problem},'' \emph{Journal of the American Statistical Association}, vol.~82, no. 397, pp. 106--111, 1987.

\bibitem{aubert_other_2006}
``\BIBforeignlanguage{en}{Other {Challenging} {Applications}},'' in \emph{\BIBforeignlanguage{en}{Mathematical {Problems} in {Image} {Processing}: {Partial} {Differential} {Equations} and the {Calculus} of {Variations}}}, ser. Applied {Mathematical} {Sciences}, G.~Aubert and P.~Kornprobst, Eds.\hskip 1em plus 0.5em minus 0.4em\relax New York, NY: Springer, 2006, pp. 213--305.

\bibitem{bergh_sample_2015}
D.~Bergh, ``\BIBforeignlanguage{en}{Sample {Size} and {Chi}-{Squared} {Test} of {Fit}—{A} {Comparison} {Between} a {Random} {Sample} {Approach} and a {Chi}-{Square} {Value} {Adjustment} {Method} {Using} {Swedish} {Adolescent} {Data}},'' in \emph{\BIBforeignlanguage{en}{Pacific {Rim} {Objective} {Measurement} {Symposium} ({PROMS}) 2014 {Conference} {Proceedings}}}, Q.~Zhang and H.~Yang, Eds.\hskip 1em plus 0.5em minus 0.4em\relax Berlin, Heidelberg: Springer, 2015, pp. 197--211.

\bibitem{lin_research_2013}
M.~Lin, H.~C. Lucas, and G.~Shmueli, ``Research {Commentary}—{Too} {Big} to {Fail}: {Large} {Samples} and the p-{Value} {Problem},'' \emph{Information Systems Research}, vol.~24, no.~4, pp. 906--917, Dec. 2013.

\bibitem{colquhoun_reproducibility_2017}
D.~Colquhoun, ``The reproducibility of research and the misinterpretation of p-values,'' \emph{Royal Society Open Science}, vol.~4, no.~12, p. 171085, Dec. 2017.

\bibitem{lakens_equivalence_2017}
D.~Lakens, ``\BIBforeignlanguage{en}{Equivalence {Tests}: {A} {Practical} {Primer} for t {Tests}, {Correlations}, and {Meta}-{Analyses}},'' \emph{\BIBforeignlanguage{en}{Social Psychological and Personality Science}}, vol.~8, no.~4, pp. 355--362, May 2017.

\bibitem{halsey_reign_2019}
L.~G. Halsey, ``The reign of the p-value is over: what alternative analyses could we employ to fill the power vacuum?'' \emph{Biology Letters}, vol.~15, no.~5, p. 20190174, May 2019.

\bibitem{boyle_method_1986}
J.~P. Boyle and R.~L. Dykstra, ``\BIBforeignlanguage{en}{A {Method} for {Finding} {Projections} onto the {Intersection} of {Convex} {Sets} in {Hilbert} {Spaces}},'' in \emph{\BIBforeignlanguage{en}{Advances in {Order} {Restricted} {Statistical} {Inference}}}, R.~Dykstra, T.~Robertson, and F.~T. Wright, Eds.\hskip 1em plus 0.5em minus 0.4em\relax New York, NY: Springer, 1986, pp. 28--47.

\bibitem{gilitschenski_direct_2014}
I.~Gilitschenski and U.~D. Hanebeck, ``A direct method for checking overlap of two hyperellipsoids,'' in \emph{2014 {Sensor} {Data} {Fusion}: {Trends}, {Solutions}, {Applications} ({SDF})}, Oct. 2014, pp. 1--6.

\bibitem{rabiei_intersection_2021}
N.~Rabiei and E.~G. Saleeby, ``\BIBforeignlanguage{en}{On intersection volumes of confidence hyper-ellipsoids and two geometric {Monte} {Carlo} methods},'' \emph{\BIBforeignlanguage{en}{Monte Carlo Methods and Applications}}, vol.~27, no.~2, pp. 153--167, Jun. 2021.

\bibitem{alnaes_fenics_2015}
M.~Alnæs \emph{et~al.}, ``\BIBforeignlanguage{en}{The {FEniCS} {Project} {Version} 1.5},'' \emph{\BIBforeignlanguage{en}{Archive of Numerical Software}}, vol.~3, no. 100, Dec. 2015, number: 100.

\bibitem{villa_hippylib_2018}
U.~Villa, N.~Petra, and O.~Ghattas, ``\BIBforeignlanguage{en}{{hIPPYlib}: {An} {Extensible} {Software} {Framework} for {Large}-{Scale} {Inverse} {Problems}},'' \emph{\BIBforeignlanguage{en}{Journal of Open Source Software}}, vol.~3, no.~30, p. 940, Oct. 2018.

\bibitem{the_cgal_project_cgal_2024}
{The CGAL Project}, \emph{{CGAL} {User} and {Reference} {Manual}}, 6th~ed.\hskip 1em plus 0.5em minus 0.4em\relax CGAL Editorial Board, 2024.

\bibitem{morozov_criteria_1984}
V.~A. Morozov, ``Criteria for {Selection} of {Regularization} {Parameter},'' in \emph{Methods for {Solving} {Incorrectly} {Posed} {Problems}}, V.~A. Morozov, Ed.\hskip 1em plus 0.5em minus 0.4em\relax New York, NY: Springer, 1984, pp. 32--64.

\bibitem{borzi_computational_2011}
A.~Borzì and V.~Schulz, \emph{\BIBforeignlanguage{en}{Computational {Optimization} of {Systems} {Governed} by {Partial} {Differential} {Equations}}}.\hskip 1em plus 0.5em minus 0.4em\relax Society for Industrial and Applied Mathematics, Jan. 2011.

\bibitem{spantini_optimal_2015}
A.~Spantini \emph{et~al.}, ``Optimal {Low}-rank {Approximations} of {Bayesian} {Linear} {Inverse} {Problems},'' \emph{SIAM Journal on Scientific Computing}, vol.~37, no.~6, pp. A2451--A2487, Jan. 2015.

\bibitem{saibaba_randomized_2016}
A.~K. Saibaba, J.~Lee, and P.~K. Kitanidis, ``\BIBforeignlanguage{en}{Randomized algorithms for generalized {Hermitian} eigenvalue problems with application to computing {Karhunen}–{Loève} expansion},'' \emph{\BIBforeignlanguage{en}{Numerical Linear Algebra with Applications}}, vol.~23, no.~2, pp. 314--339, 2016.

\bibitem{halko_finding_2011}
N.~Halko, P.~G. Martinsson, and J.~A. Tropp, ``Finding {Structure} with {Randomness}: {Probabilistic} {Algorithms} for {Constructing} {Approximate} {Matrix} {Decompositions},'' \emph{SIAM Review}, vol.~53, no.~2, pp. 217--288, Jan. 2011.

\bibitem{mittal_high-resolution_2021}
N.~Mittal and S.~J. Han, ``\BIBforeignlanguage{en}{High-{Resolution}, {Highly}-{Integrated} {Traction} {Force} {Microscopy} {Software}},'' \emph{\BIBforeignlanguage{en}{Current Protocols}}, vol.~1, no.~9, p. e233, 2021.

\bibitem{mittal_myosin-independent_2024}
N.~Mittal \emph{et~al.}, ``\BIBforeignlanguage{en}{Myosin-independent stiffness sensing by fibroblasts is regulated by the viscoelasticity of flowing actin},'' \emph{\BIBforeignlanguage{en}{Communications Materials}}, vol.~5, no.~1, pp. 1--19, Jan. 2024.

\bibitem{aguilar-cuenca_nuevos_2017}
R.~Aguilar-Cuenca, ``Nuevos mecanismos de regulación de las funciones celulares de la proteína contráctil miosina {II} no muscular,'' Ph.D. dissertation, Universidad Autónoma de Madrid, 2017.

\bibitem{kass_snakes_1988}
M.~Kass, A.~Witkin, and D.~Terzopoulos, ``\BIBforeignlanguage{en}{Snakes: {Active} contour models},'' \emph{\BIBforeignlanguage{en}{International Journal of Computer Vision}}, vol.~1, no.~4, pp. 321--331, Jan. 1988.

\bibitem{manich_protocol_2020}
M.~Manich, A.~Boquet-Pujadas, S.~Dallongeville, N.~Guillen, and J.-C. Olivo-Marin, ``\BIBforeignlanguage{en}{A {Protocol} to {Quantify} {Cellular} {Morphodynamics}: {From} {Cell} {Labelling} to {Automatic} {Image} {Analysis}},'' in \emph{\BIBforeignlanguage{en}{Eukaryome {Impact} on {Human} {Intestine} {Homeostasis} and {Mucosal} {Immunology}}}.\hskip 1em plus 0.5em minus 0.4em\relax Cham: Springer International Publishing, 2020, pp. 351--367.

\bibitem{beck_fast_2009}
A.~Beck and M.~Teboulle, ``A {Fast} {Iterative} {Shrinkage}-{Thresholding} {Algorithm} for {Linear} {Inverse} {Problems},'' \emph{SIAM Journal on Imaging Sciences}, vol.~2, no.~1, pp. 183--202, Jan. 2009.

\bibitem{lefkimmiatis_hessian_2013}
S.~Lefkimmiatis, J.~P. Ward, and M.~Unser, ``Hessian {Schatten}-{Norm} {Regularization} for {Linear} {Inverse} {Problems},'' \emph{IEEE Transactions on Image Processing}, vol.~22, no.~5, pp. 1873--1888, May 2013.

\bibitem{boquet-pujadas_sensitivity-aware_2024}
A.~Boquet-Pujadas, P.~d.~A. Pla, and M.~Unser, ``Sensitivity-{Aware} {Density} {Estimation} in {Multiple} {Dimensions},'' \emph{IEEE Transactions on Pattern Analysis and Machine Intelligence}, pp. 1--16, 2024.

\bibitem{pourya_box-spline_2024}
M.~Pourya, A.~Boquet-Pujadas, and M.~Unser, ``A {Box}-{Spline} {Framework} for {Inverse} {Problems} {With} {Continuous}-{Domain} {Sparsity} {Constraints},'' \emph{IEEE Transactions on Computational Imaging}, vol.~10, pp. 790--805, 2024.

\bibitem{javanmard_confidence_2013}
A.~Javanmard and A.~Montanari, ``Confidence {Intervals} and {Hypothesis} {Testing} for {High}-{Dimensional} {Statistical} {Models},'' in \emph{Advances in {Neural} {Information} {Processing} {Systems}}, vol.~26.\hskip 1em plus 0.5em minus 0.4em\relax Curran Associates, Inc., 2013.

\bibitem{bobkov_concentration_2011}
S.~Bobkov and M.~Madiman, ``Concentration of the {Information} in {Data} with {Log}-{Concave} {Distributions},'' \emph{The Annals of Probability}, vol.~39, no.~4, pp. 1528--1543, 2011.

\bibitem{bauschke_projection_1996}
H.~H. Bauschke and J.~M. Borwein, ``On {Projection} {Algorithms} for {Solving} {Convex} {Feasibility} {Problems},'' \emph{SIAM Review}, vol.~38, no.~3, pp. 367--426, Sep. 1996.

\bibitem{good_bayesnon-bayes_1992}
I.~J. Good, ``The {Bayes}/{Non}-{Bayes} {Compromise}: {A} {Brief} {Review},'' \emph{Journal of the American Statistical Association}, vol.~87, no. 419, pp. 597--606, Sep. 1992.

\end{thebibliography}

\end{document}